\documentclass{aa}  
\usepackage[utf8]{inputenc}
\usepackage[T1]{fontenc}
\usepackage{color}
\usepackage{multirow}
\usepackage{amsmath}
\usepackage{graphicx}
\usepackage{txfonts}
\usepackage[hidelinks=true]{hyperref}
\usepackage{xcolor}

\usepackage{placeins}

\hypersetup{
    colorlinks,
   linkcolor={red!50!black},
   citecolor={blue!50!black},
  urlcolor={blue!80!black}
}

\usepackage{gensymb}
\usepackage{wasysym}
\usepackage[flushleft]{threeparttable}

\usepackage{natbib}

\usepackage{footmisc}

\begin{document} 
	
\title{Simulating the Environment Around Planet-Hosting Stars}
\subtitle{I. Coronal Structure}

\author{J. D. Alvarado-G\'omez\inst{1,2}, G. A. J. Hussain\inst{1,3}, O. Cohen\inst{4}, J. J. Drake\inst{4}, C. Garraffo\inst{4}, J. Grunhut\inst{1} 
         	%\inst{1} 
	\and	
	 T. I. Gombosi\inst{5}%\fnmsep\thanks{Just to show the usage
          %of the elements in the author field}
          }
          \institute{\inst{1} European Southern Observatory,
              Karl-Schwarzschild-Str. 2, 85748 Garching bei M\"unchen, Germany\\
              \email{jalvarad@eso.org} \\
              \inst{2} Universit\"ats-Sternwarte M\"unchen, Ludwig-Maximilians-Universit\"at, Scheinerstr.~1, 81679 M\"unchen, Germany \\
               \inst{3} Institut de Recherche en Astrophysique et Plan\'etologie, Universit\'e de Toulouse, UPS-OMP, F-31400 Toulouse, France \\
              \inst{4} Harvard-Smithsonian Center for Astrophysics, 60 Garden Street, Cambridge, MA 02138, USA\\
              \inst{5} Center for Space Environment Modeling, University of Michigan, 2455 Hayward St., Ann Arbor, MI 48109, USA	
             }
             
   \date{Received -----; accepted -----}

% \abstract{}{}{}{}{} 
% 5 {} token are mandatory
 
\abstract{We present the results of a detailed numerical simulation of the circumstellar environment around three exoplanet-hosting stars. A state-of-the-art global magnetohydrodynamic (MHD) model is considered, including Alfv\'en wave dissipation as a self-consistent coronal heating mechanism. This paper contains the description of the numerical set-up, evaluation procedure, and the simulated coronal structure of each system (HD 1237, HD 22049 and HD 147513). The simulations are driven by surface magnetic field maps, recovered with the observational technique of Zeeman Doppler Imaging (ZDI). A detailed comparison of the simulations is performed, where two different implementations of this mapping routine are used to generate the surface field distributions. Quantitative and qualitative descriptions of the coronae of these systems are presented, including synthetic high-energy emission maps in the Extreme Ultra-Violet (EUV) and Soft X-rays (SXR) ranges. Using the simulation results, we are able to recover similar trends as in previous observational studies, including the relation between the magnetic flux and the coronal X-ray emission. Furthermore, for HD 1237 we estimate the rotational modulation of the high-energy emission due to the various coronal features developed in the simulation. We obtain variations, during a single stellar rotation cycle, up to 15\% for the EUV and SXR ranges. The results presented here will be used, in a follow-up paper, to self-consistently simulate the stellar winds and inner astrospheres of these systems.}   


\titlerunning{Environment around Planet-Hosting Stars - I. Coronal Structure}
\authorrunning{Alvarado-G\'omez et al.}
\maketitle
%
%________________________________________________________________

\section{Introduction}\label{sec_intro}

\noindent Analogous to the 11-year Solar activity cycle, a large fraction of late-type stars ($\sim 60\%$) show chromospheric activity cycles, with periods ranging from 2.5 to 25 years \citep{1995ApJ...438..269B}. For a very limited number of these systems, including binaries, the coronal X-ray counterparts of these activity cycles have also been identified (e.g. \citealt{2008A&A...490.1121F,2012A&A...543A..84R}). These periodic signatures appear as a result of the magnetic cycle of the star. In the case of the Sun, this is completed every 22 years over which the polarity of the large-scale magnetic field is reversed twice \citep{2010LRSP....7....1H}. These elements, the cyclic properties of the activity and magnetic field, constitute a major benchmark for any dynamo mechanism proposed for the magnetic field generation \citep{2014ARA&A..52..251C}.

%Apart from their direct connection with the activity cycles, magnetic fields strongly influence different aspects of the stellar structure and evolution. It is believed that they play a major role in the coronal heating problem in the Sun and other late type stars \citep{2012RSPTA.370.3217P}. In addition, the development of magnetically-driven stellar winds is crucial to understand  the evolution of cool stars in the early main sequence. G to K type stars tend to rotate rapidly on the Zero Age Main Sequence; braking torques exerted by winds (or outflows) cause them to spin down, losing most of their angular momentum within the first 500 Myr on the main sequence \citep{2010ApJ...721..675B}.

Recent developments in instrumentation and observational techniques have opened a new window for stellar magnetic field studies across the HR diagram (see \citealt{2009ARA&A..47..333D}). In particular, the large-scale surface magnetic field topology in stars different from the Sun can be retrieved using the technique of Zeeman Doppler Imaging (ZDI, \citealt{1989A&A...225..456S, 1991A&A...250..463B, 1997A&A...326.1135D, 2002A&A...381..736P, 2009MNRAS.398..189H,2010A&A...513A..13K}). Several studies have shown the robustness of this procedure, successfully recovering the field distribution on the surfaces of Sun-like stars, over a wide range of activity levels (e.g. \citealt{2008MNRAS.385.1179D, 2008MNRAS.388...80P, 2015A&A...582A..38A, 2016A&A...585A..77H}). Long-term ZDI monitoring of particular Sun-like targets have shown different time-scales of variability in the large-scale magnetic field. This includes fast and complex evolution without polarity reversals (e.g. HN Peg, \citealt{2015A&A...573A..17B}), erratic polarity changes (e.g. $\xi$ Boo, \citealt{2012A&A...540A.138M}) and hints of magnetic cycles with single (e.g. HD 190771, \citealt{2009A&A...508L...9P}), and double (e.g. $\tau$ Boo, \citealt{2009MNRAS.398.1383F}) polarity reversals in a time-scale of 1-2 years.

Furthermore, ZDI maps have proven to be very useful in other aspects of cool stellar systems research. Applications cover magnetic activity modelling for radial velocity jitter corrections \citep{2014MNRAS.444.3220D}, transit variability and bow-shocks \citep{2013MNRAS.436.2179L}, coronal X-ray emission \citep{2010MNRAS.404..101J, 2011MNRAS.410.2472A, 2014MNRAS.439.2122L} and mass loss rates in connection with stellar winds \citep{2010ApJ...721...80C, 2011MNRAS.412..351V}. 

%Average magnetic energy density (UBr = Br^2/8\pi) over the visible surface (limited by the inclination)
%			ZDI		SH-ZDI
%HD 1237 	7.80		25.94	
%HD 22049	3.70		?
%HD147513	--		8.29

In the case of planet-hosting systems, ZDI-based studies have tended to focus on close-in exoplanet environments by applying detailed global three-dimensional magnetohydrodynamic (MHD) models, originally developed for the solar system (BATS-R-US code, \citealt{1999JCoPh.154..284P}). This numerical treatment includes all the relevant physics for calculating a stellar corona/wind model, using the surface magnetic field maps as driver of a steady-state solution for each system. Within the MHD regime, two main approaches have been considered: an ad-hoc thermally-driven polytropic stellar wind (i.e., $P \propto \rho^\gamma$, with $\gamma$ as the polytropic index, \citealt{2011ApJ...738..166C,2012MNRAS.423.3285V,2015MNRAS.449.4117V}), and a more recent description, with Alfv\'en wave turbulence dissipation as a self-consistent driver of the coronal heating and the stellar wind acceleration in the model \citep{2014ApJ...790...57C}. This last scheme is grounded on strong observational evidence that Alfv\'en waves, of sufficient strength to drive the solar wind, permeate the solar chromosphere \citep{2007Sci...318.1574D,2011Natur.475..477M}. Additionally, this numerical approach has been extensively validated against STEREO/EUVI and SDO/AIA measurements (see \citealt{2014ApJ...782...81V}). The models presented in this paper are based on this latest treatment of the heating and energy transfer in the corona.

In this work we present the results of a detailed numerical simulation of the circumstellar environment around three late-type exoplanet-hosts (HD 1237, HD 22049 and HD 147513), using a 3D MHD model. This first article contains the results of the simulated coronal structure, while the wind and inner astrosphere domains will be presented in a follow-up paper. The simulations are driven by the radial component of the large-scale surface magnetic field in these stars, which have been recovered using two different implementations of ZDI (Sect. \ref{sec_stars}). All three systems have similar coronal (X-ray) activity levels (see table \ref{table_1}). While these are more active than the Sun they would be classified as moderately active stars and well below the X-ray/activity saturation level. 
%In the particular case of HD 1237, we also investigate the interactions of the magnetized stellar wind with the exoplanet of this system, assuming a Jupiter-like magnetosphere around it. 
A description of the numerical set-up is provided in section \ref{sec_num}, and the results are presented in section \ref{sec_results}. Section \ref{sec_discussion} contains a discussion in the context of other studies and the conclusions of our work are summarized in section \ref{sec_conclusions}. 

\section{Large-Scale Magnetic Field Maps}\label{sec_stars}

HD 1237, HD 147513 and HD 22049 are cool main sequence stars (G8, G5 and K2 respectively) with relatively slow rotation rates ($P_{\rm rot}\sim$\,$7-12$ days). Each of these systems host a Jupiter-mass planet ($M_{\rm p}\sin i > M_{\jupiter}$), with orbital separations comparable to the solar system planets \citep{2000ApJ...544L.145H,2001A&A...375..205N,2004A&A...415..391M,2006AJ....132.2206B}. Table \ref{table_1} contains a summary of the relevant astrophysical parameters for each system, taken from various observational studies. 
 
%Two G- (HD 1237 and HD 147513), and one K-dwarf (HD 22049), are the targets considered in this study. Each of these systems hosts a Jupiter-mass planet ($M_{\rm p}\sin i > M_{\jupiter}$), with orbital separations comparable to the solar system planets \citep{2000ApJ...544L.145H,2001A&A...375..205N,2004A&A...415..391M,2006AJ....132.2206B}. Table \ref{table_1} contains a summary of the relevant astrophysical parameters for each system. 

%Table from next section.
\begin{table*}
\caption{Planet-hosting systems and their observational properties.}             
\label{table_1}      
\centering
{\small          
\begin{tabular}{l c c c c c c c c c c c | c c}    
\hline\hline

Star ID & S. Type & $T_{\rm eff}$ & $R_*$ & $M_{*}$ & $P_{\rm rot}$ & $i$ & Age &  \multicolumn{2}{c}{Activity} & $M_{\rm p}\sin i$ & $a$ & \multicolumn{2}{c}{$\left<\varepsilon_{B\,r}\right>$}\\
 & & [K] & [$R_{\odot}$] & [$M_{\odot}$] & [days] & [$^{\circ}$] & [Gyr] & $\log(R^{\prime}_{\rm HK})$ & $\log(L_{\rm X})$ & [$M_{\jupiter}$] & [AU]\vspace{2pt} & ZDI & SH-ZDI\\
\hline                    
HD 1237\,\tablefootmark{a} & G8V & 5572 & 0.86 & 1.00 & 7.00 & $\sim$\,50 & $\sim$0.88 & $-4.38$ & $29.02$ & 3.37 & 0.49 & $4.65$ & $30.77$\\  
HD 22049\,\tablefootmark{b} & K2V & 5146 & 0.74 & 0.86 & 11.68 & $\sim$\,45 & $\sim$0.44 & $-4.47$ & $28.22$ & 1.55 & 3.39 & $2.32$ & $ 30.66 $ \\  
HD 147513\,\tablefootmark{c} & G5V & 5930 & 0.98 & 1.07 & 10.00 & $\sim$\,20 & $\sim$0.45 & $-4.64$ & $28.92$ & 1.21 & 1.32 & $-$\tablefootmark{$\dagger$} & $6.21$\\  
\hline                  
\end{tabular}}
\tablefoot{The values listed in columns $1 - 12$ are taken from previous studies of each system and references therein: \tablefoottext{a}{\cite{2001A&A...375..205N,2015A&A...582A..38A}} \tablefoottext{b}{\cite{1993ApJ...412..797D,2000ApJ...544L.145H, 2006AJ....132.2206B, 2014A&A...569A..79J}} and \tablefoottext{c}{\cite{2004A&A...415..391M,2016A&A...585A..77H}.} The last two columns contain the (radial) magnetic energy density, $\varepsilon_{\,B\,r} = B_{\rm r}^2 / 8\pi$, averaged over the visible surface of the star, and estimated from the standard ZDI and the Spherical Harmonics implementation (SH-ZDI).\\ \tablefoottext{$\dagger$}{Due to the low inclination and simple field geometry, the standard ZDI reconstruction was not possible in this case (see \citealt{1991A&A...250..463B}).}
}
\end{table*}

Previous works have recovered the large-scale magnetic field on the surfaces of these stars, by applying ZDI to time-series of circularly polarised spectra (\citealt{2014A&A...569A..79J,2015A&A...582A..38A,2016A&A...585A..77H}). For the stars included in this work, this has been done with the spectropolarimeter NARVAL at the Telescope Bernard Lyot \citep{2003EAS.....9..105A}, and the polarimetric mode \citep{2011Msngr.143....7P} of the HARPS echelle spectrograph \citep{2003Msngr.114...20M} on the ESO 3.6\,m telescope at La Silla Observatory. For consistency, the ZDI maps included in the simulations have been reconstructed using data from the same instrument/telescope\footnote[2]{Therefore, for HD 22049 ($\epsilon$~Eridani) we only consider the January 2010 dataset (see \citealt{2011Msngr.143....7P, 2014A&A...569A..79J}).} (i.e. HARPSpol).  

For the magnetic field mapping procedure, we considered two different approaches; the classic ZDI reconstruction, in which each component of the magnetic field vector is decomposed in a series of independent magnetic-image pixels \citep{1991A&A...250..463B, 1997A&A...326.1135D}, and the spherical harmonics decomposition (SH-ZDI) where the field is described by the sum of a potential and a toroidal component, and each component is expanded in a spherical-harmonics basis (see \citealt{2001MNRAS.322..681H, 2006MNRAS.370..629D}). Both procedures are equivalent, leading to very similar field distributions and associated fits to the spectro-polarimetric data. However, as described by \citet{1991A&A...250..463B}, ZDI is not able to properly recover very simple field geometries (e.g. dipoles), and is more suitable for complex (spotted) magnetic distributions. This limitation is removed in the SH-ZDI implementation. Both procedures are restricted by the inclination angle of the star and therefore, a fraction of the surface field that cannot be observed, is not recovered in the maps. To correct for this effect, previous numerical studies have completed the field distribution by a reflection of the ZDI map across the equatorial plane (e.g. \citealt{2010ApJ...721...80C}). More recently, \citet{2012MNRAS.423.3285V} have included complete symmetric/antisymmetric SH-ZDI maps to show that the map incompleteness has a minor impact over their simulation results. However, for the simulations performed here, which include the latest implementation of BATS-R-US, this may not be the case. A larger impact may be expected on the overall coronal structure, as the mechanism for the coronal heating and the wind acceleration is directly related to the field strength and topology (e.g. Alfv\'en waves, see \citealt{2014ApJ...782...81V}). 

\begin{figure}[ht]
\centering %  left, bottom, right and top
\includegraphics[trim=0.4cm 0.4cm 0.25cm 0.7cm, clip=true, width=\hsize/2]{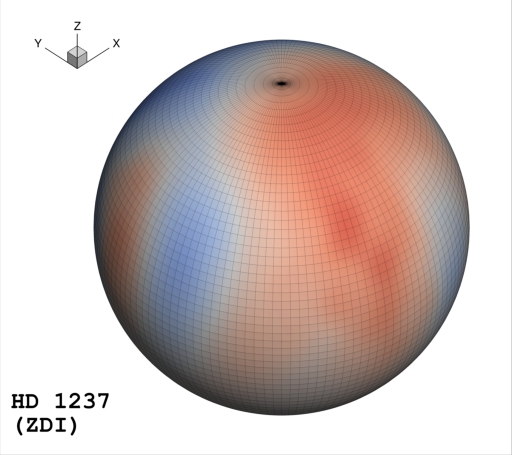}\includegraphics[trim=0.4cm 0.4cm 0.25cm 0.7cm, clip=true, width=\hsize/2]{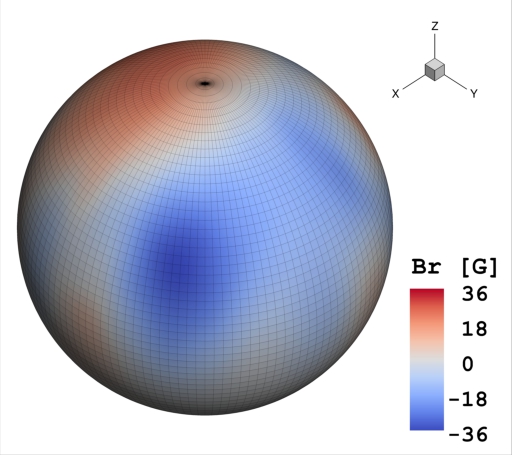}
\includegraphics[trim=0.4cm 0.0cm 0.25cm 1.2cm, clip=true, width=\hsize/2]{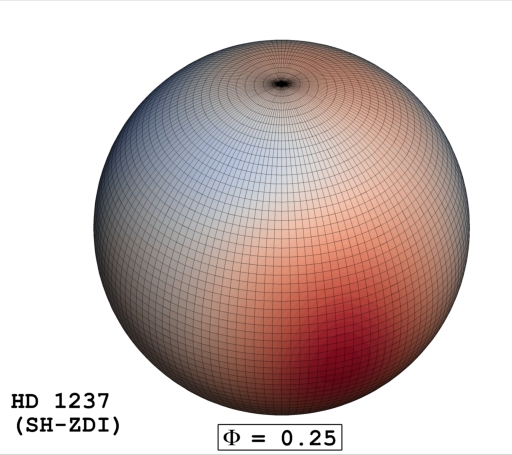}\includegraphics[trim=0.4cm 0.0cm 0.25cm 1.2cm, clip=true, width=\hsize/2]{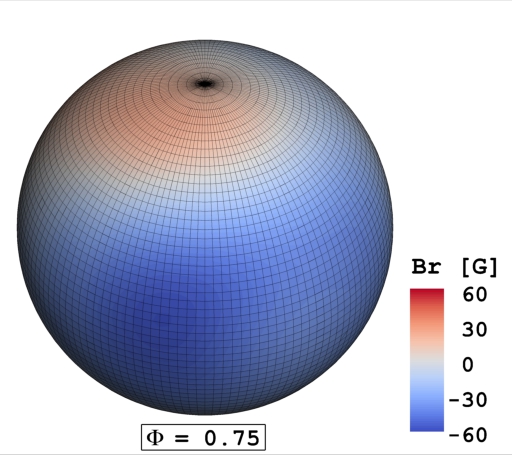}
\caption{Surface radial magnetic field maps of HD 1237. A comparison between the standard ZDI (top) and the SH-ZDI (bottom) is presented. The colour scale indicates the polarity and the field strength in Gauss (G). Note the difference in the magnetic field range for each case. The stellar inclination angle ($i = 50\,^{\circ}$) is used for the visualizations.}
\label{fig_1}
\end{figure}

\begin{figure}[!ht]
\centering %  left, bottom, right and top
\includegraphics[trim=0.4cm 0.4cm 0.25cm 0.7cm, clip=true, width=\hsize/2]{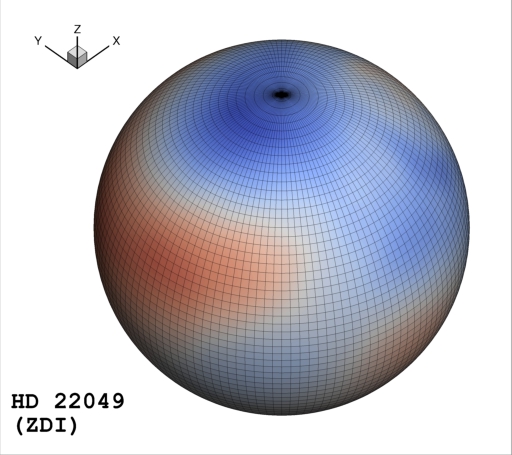}\includegraphics[trim=0.4cm 0.4cm 0.25cm 0.7cm, clip=true, width=\hsize/2]{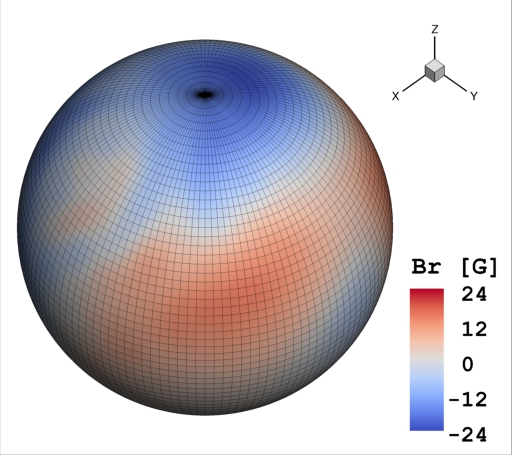}
\includegraphics[trim=0.4cm 0.0cm 0.25cm 1.2cm, clip=true, width=\hsize/2]{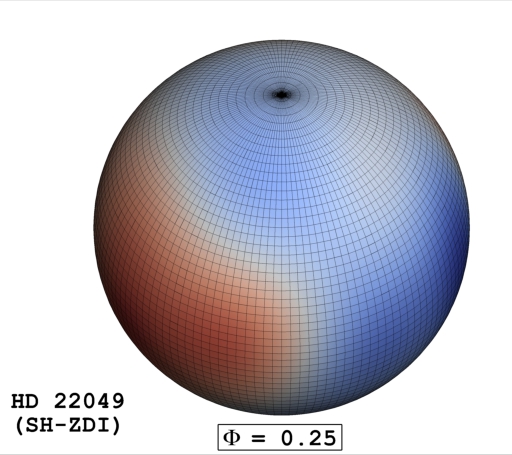}\includegraphics[trim=0.4cm 0.0cm 0.25cm 1.2cm, clip=true, width=\hsize/2]{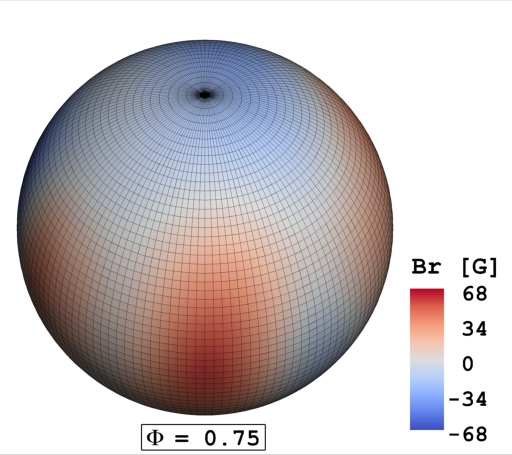}
\caption{Surface radial magnetic field maps of HD 22049. See caption of Fig. \ref{fig_1}. The stellar inclination angle ($i = 45\,^{\circ}$) is used for the visualizations.}
\label{fig_2}
\end{figure}

\begin{figure}[!ht]
\centering %  left, bottom, right and top
\includegraphics[trim=0.4cm 0.0cm 0.25cm 0.7cm, clip=true, width=\hsize/2]{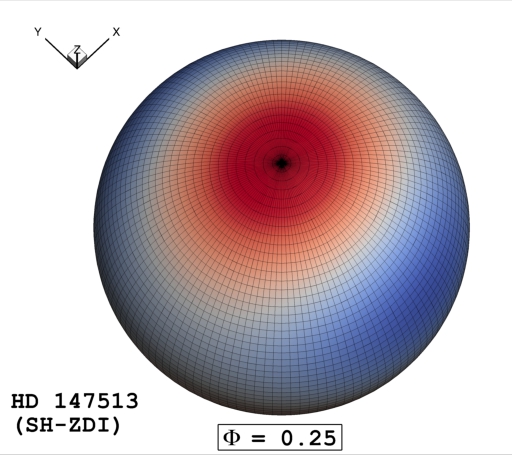}\includegraphics[trim=0.4cm 0.0cm 0.25cm 0.7cm, clip=true, width=\hsize/2]{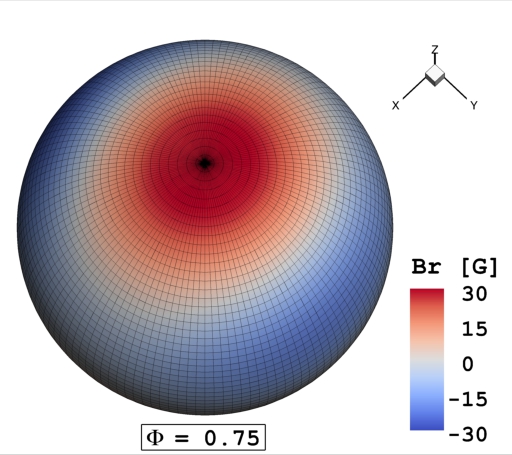}
\caption{Surface radial magnetic field maps of HD 147513 using SH-ZDI. Two rotational phases ($\Phi$) are presented. The stellar inclination angle ($i = 20\,^{\circ}$) is used for the visualizations.}
\label{fig_3}
\end{figure}

\begin{figure}[!ht]
\centering %  left, bottom, right and top
\includegraphics[trim=0.4cm 0.4cm 0.25cm 0.7cm, clip=true, width=\hsize/2]{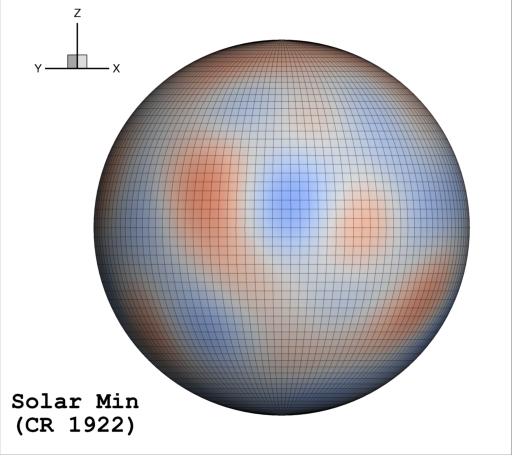}\includegraphics[trim=0.4cm 0.4cm 0.25cm 0.7cm, clip=true, width=\hsize/2]{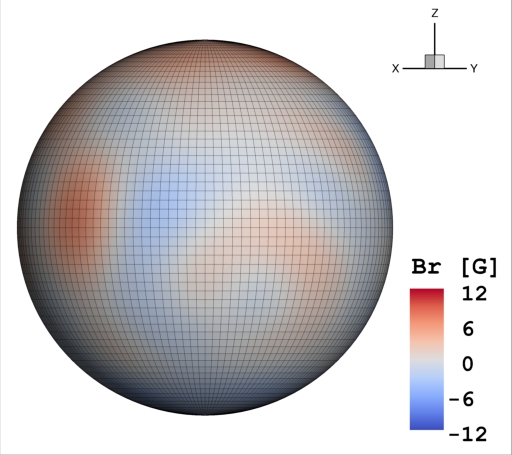}
\includegraphics[trim=0.4cm 0.0cm 0.25cm 1.2cm, clip=true, width=\hsize/2]{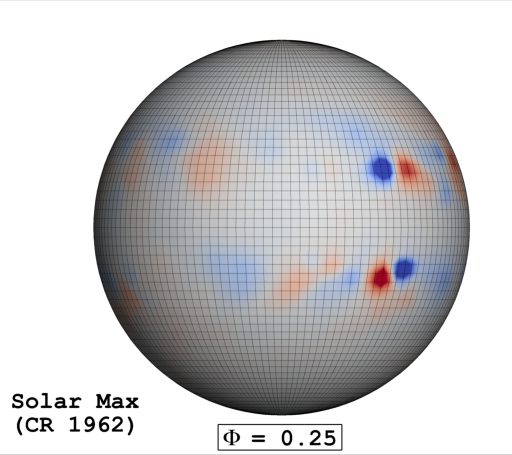}\includegraphics[trim=0.4cm 0.0cm 0.25cm 1.2cm, clip=true, width=\hsize/2]{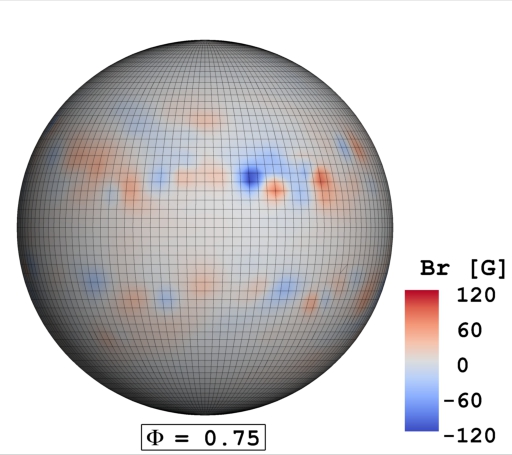}
\caption{Surface radial magnetic field maps of the Sun during activity minimum (CR 1922, top) and maximum (CR 1962, bottom) taken by SOHO/MDI. Note the difference in the magnetic field range for each case. An inclination angle $i = 90\,^{\circ}$ is used for the visualizations.}
\label{fig_4}
\end{figure}

\noindent Figures \ref{fig_1} and \ref{fig_2} show a comparison between the reconstruction procedures applied to HD 1237 and HD 22049, respectively. In general, the maps obtained using ZDI show a more complex and weaker field distribution in comparison to the SH-ZDI, where a smoother field topology is obtained. While there are similarities in the large-scale structure, discrepancies are obtained in terms of the amount of detail recovered in each case. These differences arise as a consequence of the constraints imposed for completing the SH-ZDI maps, which are all pushed to symmetric field distributions. In general, the spatial resolution of the SH-ZDI maps depend on the maximum order of the spherical harmonics expansion ($l_{\rm max}$). For each case this is selected in such a way that the lowest possible $l_{\rm max}$ value is used, while achieving a similar goodness-of-fit level (reduced $\chi^2$) as the classic ZDI reconstruction (HD 1237: $l_{\rm max} = 5$, HD 22049: $l_{\rm max} = 6$, HD147513: $l_{\rm max} = 4$). Higher values of  $l_{\rm max}$ would not alter the large-scale distribution, but introduces further small-scale field without significantly improving the goodness-of-fit. This step is particularly important for a consistent comparison, as the final recovered field strengths depend on this. All these differences have a significant impact in the coronal and wind structure, as they depend on the field coverage and the amount of magnetic energy available in each case (see Table \ref{table_1}). In the case of HD 147513 the standard ZDI reconstruction was not possible, given its low inclination angle ($i \sim 20^{\circ}$) and fairly simple large-scale topology. Therefore, for this system we only consider the SH-ZDI map presented in Fig. \ref{fig_3}, previously published by \citet{2016A&A...585A..77H}.

To evaluate our numerical results, we have performed two additional simulations taking the Sun as reference. The magnetic field distributions during solar minimum (Carrington rotation 1922, end of cycle 22), and solar maximum (Carrington rotation 1962, during cycle 23) have been considered for this purpose. The large-scale magnetic field is taken from synoptic magnetograms, generated by the Michelson Doppler Imager instrument (MDI, \citealt{1995SoPh..162..129S}) on board the Solar and Heliospheric Observatory spacecraft (SOHO, \citealt{1995SoPh..162....1D}). Figure \ref{fig_4} shows the comparison between the global magnetic field distribution for these activity epochs. During activity minimum, weak magnetic regions (a few Gauss) tend to be sparsely distributed across the entire solar surface (no preferential location for these regions is observed). Stronger small-scale magnetic fields, up to two orders of magnitude, can be found during activity maximum. In this case the dominant fields are highly concentrated in bipolar sectors (active regions) and located mainly in two latitudinal belts at $\sim \pm 30^{\circ}$. Still, weaker magnetic fields can be found along the entire solar surface.

Finally, as is shown in Figs. \ref{fig_1} to \ref{fig_4}, the numerical grid for all the input surface magnetic field distributions is the same. Therefore, the resolution of the solar coronal models was adapted to match the optimal resolution of the stellar simulations. In this way, a more consistent comparison of the results can be performed. The surface grid resolution ($\sim$\,$10^{-2}$ $R_{*}$) is sufficient to resolve entirely the magnetic structures on the stellar ZDI/SH-ZDI maps. However, in the solar case the internal structure of the active regions and the small-scale structures are not resolved. The impact of this limited resolution in magnetic field maps for solar simulations has been investigated previously by \citet{2013ApJ...764...32G}. They found that the structure of the stellar wind is less sensitive to this factor than the coronal structure and associated emission (e.g. EUV and X-rays). This will be explored in more detail in the evaluation procedure, presented in Sect. \ref{SC_Calibration}.

\section{3D MHD Numerical Simulation}\label{sec_num}

\noindent The numerical simulations presented here are performed using the three-dimensional MHD code BATS-R-US \citep{1999JCoPh.154..284P} as part of the Space Weather Modeling Framework (SWMF, \citealt{2012JCoPh.231..870T}). As discussed previously by \citet{2014ApJ...790...57C}, the SWMF encompasses a collection of physics-based models for different regimes in solar and space physics. These can be considered individually or can be coupled together to provide a more realistic description of the phenomenon or domain of interest. For the systems considered here, we have included and coupled two overlapping domains to obtain a robust combined solution. The results presented in this paper correspond to the stellar corona domain (SC module). A follow-up study will contain the wind and the inner astrosphere description (IH module). 
%In the case of HD 1237, we have also coupled a global magnetospheric model (GM module) to investigate the conditions experienced by the exoplanet in its orbit. 
The solution for each domain is obtained using the most up-to-date version of the SWMF modules\footnote[2]{Code version 2.4}. 

The stellar corona domain extends from the base of the chromosphere ($\sim$\,$1\,R_*$) up to $30\,R_*$. A three-dimensional potential field extrapolation, above the stellar surface, is used as the initial condition. This initial extrapolation is performed based on the photospheric radial magnetic field of the star (e.g. ZDI maps, Sect. \ref{sec_stars}). In addition to the surface magnetic field distribution, this module requires information about the chromospheric base density, $n_0$, and temperature, $T_0$, as well as the stellar mass, $M_*$, radius, $R_*$ and rotation period, $P_{\rm rot}$. This differs from previous ZDI-driven numerical studies, where these thermodynamic boundary conditions are set to coronal values and therefore, not self-consistently obtained in the simulations \citep{2011ApJ...738..166C,2012MNRAS.423.3285V,2015MNRAS.449.4117V}. 

For the stars considered here, we assumed solar values for the chromospheric base density ($n_0 = 2.0\,\times\,10^{16}$ m$^{-3}$), and temperature ($T_0 = 5.0\,\times\,10^4$ K). This is justified from the fact that these systems, while more active than the Sun, are still within the X-ray un-saturated regime and therefore, the physical assumptions behind the coronal structure and the solar wind acceleration in the model are more likely to hold. This assumption permits a consistent comparison with the solar case and between the systems considered. The remaining initial required parameters for each star are listed in Table \ref{table_1}. For the solar runs we use the sidereal rotation rate of 25.38 days (Carrington rotation). 

We use a non-uniform spherical grid, dynamically refined at the locations of magnetic field inversion, which provides a maximum resolution of $\sim$\,$10^{-3}\,R_*$. The numerical simulation evolves until a steady-state solution is achieved. Coronal heating and stellar wind acceleration due to Alfv\'en wave turbulence dissipation are self-consistently calculated, taking into account electron heat conduction and radiative cooling effects. For further details the reader is referred to \citet{2013ApJ...764...23S} and \citet{2014ApJ...782...81V}.  From this final solution, all the physical properties, such as number density, $n$, plasma temperature, $T$, velocity, $\mathbf{u}$ and magnetic field, $\mathbf{B}$, can be extracted. We present the simulation results in the following section.

\section{Results}\label{sec_results}

\noindent We perform a detailed evaluation of the solution sets for the solar minimum and maximum cases in Sect. \ref{SC_Calibration}. Sections \ref{SC_HD1237} to \ref{SC_HD147513} contain the simulation results of the coronal structure for the stars considered. In each case we present the distribution of the thermodynamic conditions ($n$, $T$), as well as the magnetic energy density ($\varepsilon_{\,B\,r}$), associated with the radial field. A common colour scale is adopted for all stars to facilitate comparison\footnote[3]{Except in the magnetic energy density distribution for solar minimum case (Fig. \ref{fig_A1}), where the range is decreased by a factor of 10.}. 

In addition, synthetic coronal emission maps are generated at SXR and EUV wavelengths. This is done by integrating the square of the plasma density times the emissivity response function of a particular instrument, along the line-of-sight towards the observer. In the SXR range we consider the specific response function of the AlMg filter of the SXT/Yohkoh instrument, to synthesise images in the 2 to 30\,\AA\, range (0.25 -- 4.0 keV, red images). For the EUV range sensitivity tables of the EIT/SOHO instrument are used, leading to narrow-band images centred at the Fe IX/X 171\,\AA\, (blue), Fe XII 195\,\AA \, (green), and Fe XV 284\,\AA\, (yellow) lines. The coronal emission at these wavelengths has been extensively studied in the solar context, serving also to calibrate the results from the SWMF in various works (see \citealt{2013ApJ...764...32G, 2014ApJ...782...81V}). This procedure also allows the direct comparison of the synthetic images, generated for different stars. For HD 1237 and HD 22049 we additionally compare the results driven by the different maps of the large-scale magnetic field (Sect. \ref{sec_stars}). 

%In addition, synthetic coronal emission maps are generated at SXR (red) and EUV (blue, green and yellow) wavelengths. The SXR images are synthesised in the 2 to 30\,\AA\, range (0.25 -- 4.0 keV). The EUV images correspond to 171\,\AA\, (blue), 195\,\AA \, (green), and 284\,\AA\, (yellow). These ranges have been extensively studied in the solar context, by the Yohkoh (SXR) and SOHO (EUV) spacecrafts, and have been used to calibrate the results from the SWMF in various works (see \citealt{2013ApJ...764...32G, 2014ApJ...782...81V}). As the specific response functions for each wavelength/instrument are included in the calculations, it is possible to compare directly the images generated for different stars. For HD 1237 and HD 22049 we additionally compare the results driven by the different maps of the large-scale magnetic field (Sect. \ref{sec_stars}).  

%The 3D steady-state solutions of the coronal structure for the considered systems are presented in the following sub-sections.      %In the case of HD 1237, we also include the exoplanet environment described by the global magnetospheric model. 
\subsection{Evaluation of the Solar Case}\label{SC_Calibration}

%Table from next section
\begin{table*}
\caption{Evaluation results for the EUV range. The listed values correspond to averages over an entire rotation, obtained from the observations (Obs) and the simulations (Sim). The two filter wavelength-ratio (in $\AA$) used for the parameters estimation are indicated in each case.}             
\label{table_2}      
\centering          
\begin{tabular}{ l | c c | c c | c c | c c }    
\hline\hline
Parameter &  \multicolumn{2}{|c|}{Min (Obs)} & \multicolumn{2}{|c|}{Min (Sim)} & \multicolumn{2}{|c|}{Max (Obs)} &  \multicolumn{2}{|c}{Max (Sim)}\\
 & 195/171 & 284/195 & 195/171 & 284/195 & 195/171 & 284/195 & 195/171 & 284/195 \\
\hline                    
$\left<T\right>$ [$\times\,10^{6}$ K] & 1.06 & 1.77 & 1.14 & 1.63 & 1.13 & 1.79 & 1.16 & 1.66 \\
$\left<EM\right>$ [$\times\,10^{26}$ cm$^{-5}$] & 4.51 & 5.08 & 1.19 & 0.98 & 8.42 & 14.9 & 5.52 & 5.06 \\
\hline                  
\end{tabular}
\end{table*}

\noindent The simulation results for the Sun are presented in Appendix \ref{app_1}. The synthetic images provide a fairly good match to the solar observations obtained during 1977-May-07 (activity minimum, Fig. \ref{fig_A1}) and 2000-May-10 (activity maximum, Fig. \ref{fig_A2})\footnote[4]{For a quick-look comparison with the observations from various instruments during these dates, visit \url{http://helioviewer.org/}.}. The steady-state solution properly recovers the structural differences for both activity states. An open-field dominated corona appears in the solar minimum case, displaying coronal holes near the polar regions of the Sun. In turn, the solar maximum case shows mainly close-field regions across the solar disk, with almost no open field-line locations. This will have various implication for the associated solar wind structure, which will be discussed in the second paper of this study. 

%HD 1237: a = 0.49 AU, Msini = 3.37 M_Jup, P_orb = 133.7 days, e = 0.51
%HD 22049: a = 3.39 AU, Msini = 1.55 M_Jup, P_orb = 6.85 yr, e = 0.702 
%HD 147513: a = 1.32 AU, Msini = 1.12 M_Jup, P_orb = 528.4 days, e = 0.26

In general, the differences in the magnetic activity/complexity are clearly visible in the steady-state solution. As expected, the thermodynamic structure of the corona, and the associated high-energy emission, show large variation in both activity states. To evaluate the simulation results, we need to quantitatively compare the numerical solutions for the Sun to the real observations (i.e. based on the SXR/EUV data). As was discussed in Sect. \ref{sec_stars}, this is particularly important as the solar simulations presented here have been performed with limited spatial resolution (see also \citealt{2013ApJ...764...32G}). To do this, we contrast the simulation results to archival Yohkoh/SXT and SOHO/EIT data\footnote[1]{Available at the \href{http://sdac.virtualsolar.org/cgi/search}{Virtual Solar Observatory (VSO)}} covering both activity epochs (Carrington rotations 1922 and 1962). 

\noindent For the SXR range, we use the daily averages for the solar irradiance at 1 AU, described in \cite{1999JGR...10414827A}, and compute a mean value for each Carrington rotation. This leads to $1.02 \times 10^{-5}$ W m$^{-2}$ for solar minimum, and $1.21 \times 10^{-4}$ W m$^{-2}$ for solar maximum, in the 2 -- 30\,\AA\, range. In terms of SXR luminosities, these values correspond to $2.86 \times 10^{25}$ ergs s$^{-1}$ and $3.42 \times 10^{26}$ ergs s$^{-1}$, respectively. However, more recent estimates, presented by \citet{2003ApJ...593..534J}, lead to larger values in the SXR luminosities during the solar activity cycle (i.e. 10$^{26.8}$ ergs s$^{-1}$ during activity minimum, and 10$^{27.9}$  ergs s$^{-1}$ for activity maximum). From the steady-state solutions, we simulate the coronal emission in the SXR band with the aid of the Emission Measure distribution $EM(T)$ (Sect. \ref{cor_prop}), and following the procedure described in Sect. \ref{sec_flux}. This yields simulated values of $2.79 \times 10^{26}$ ergs s$^{-1}$ and $2.49 \times 10^{27}$ ergs s$^{-1}$ during activity minimum and maximum, respectively. 

A similar procedure is applied for the EUV range. Images acquired by the EIT instrument during both activity periods, are used for this purpose. We consider 3 full-disk images per day (one for each EUV channel, excluding the 304\,$\AA$ bandpass), for a total of 87 images per rotation. After the image processing, we performed temperature and $EM$ diagnostics, using the standard SolarSoftWare (SSW) routines for this specific instrument\footnote[2]{More information can be found in the \href{http://umbra.nascom.nasa.gov/eit/eit_guide/}{EIT user guide}}. This leads to a rough estimate of both parameters, given a pair (ratios) of EUV images. We use the temperature-sensitive line ratios of Fe XII 195\,$\AA$/ Fe IX/X 171\,$\AA$ and Fe XV 284\,$\AA$/ Fe XII 195\,$\AA$ (for a combined sensitivity range of 0.9 MK $< T <$ 2.2 MK). The reader is referred to \citet{1997SoPh..175..571M} for further information. As with the SXR range, we compute mean observed values of these parameters for both rotations, and compare them with simulated quantities, derived from the synthetic EUV emission maps. The obtained values are presented in the Table \ref{table_2}.

We also compared the synthetic EUV emission to archival data from the GOES-13/EUVS instrument\footnote[3]{See \url{http://www.ngdc.noaa.gov/stp/satellite/goes/}.}. These measurements span different solar activity periods in comparison to the epochs considered in the simulations (CR 1922 and CR 1962). Therefore, we interpret these quantities as nominal values for the EUV variation during minimum and maximum of activity. We consider GOES-13 data from channels A ($50 - 150\,\AA$) and B ($250 - 340\,\AA$), leading to average EUV luminosities, for activity minimum and maximum, of $\sim\,$$2 - 5 \times10^{27}$ erg s$^{-1}$ and $\sim\,$$1 \times10^{28}$ erg s$^{-1}$, respectively. The simulated coronal emission, synthesised in the same wavelength ranges, provides very good agreement to the observations, leading to $\sim$\,1.4\,$\times10^{27}$ erg s$^{-1}$ during solar minimum, and $\sim$\,1.3\,$\times10^{28}$ erg s$^{-1}$ at solar maximum. 

The results from the evaluation procedure are consistent between the EUV and SXR ranges, showing a reasonable match between the simulations and the overall structure of the solar corona for both activity periods. Good agreement is obtained for the low-temperature region (195/171 ratio), with differences below $+8\%$ in the mean temperature for both epochs. The sign indicates the relative difference between the simulation (Sim) and the observations (Obs). A similar level of agreement (with reversed sign) is achieved for the hotter component of the corona (284/195 ratio). Furthermore, the simulated SXR emission properly recovers the nominal estimates for both activity periods, with resulting values lying between the observational estimates of \cite{1999JGR...10414827A} and \citet{2003ApJ...593..534J}. In a similar manner, the fiducial EUV luminosities during minimum and maximum of activity are well recovered. However, we should note here that He II 304\,$\AA$ line tends to dominate the GOES-13 B bandpass. This line is overly strong compared with expectations based on collisional excitation (e.g. \citealt{1975MNRAS.170..429J, 2004ApJ...606.1239P}), and therefore our model spectrum is expected to significantly under-predict the observed flux. That we obtain reasonably good agreement is likely a result of our emission measure distribution being too high at transition region temperatures (see Sec. \ref{cor_prop}, Fig. \ref{fig_11}). In contrast, larger discrepancies are found for the $EM$ distribution (over the sensitivity range of the EIT filters) for both coronal components. During activity minimum, differences up to factors of $-3.8$ and $-5.2$ appear for the low- and high-temperature corona, respectively. Slightly smaller difference factors prevail during activity maximum for both components, reaching $-1.5$ and $-3.0$ respectively.  

Some of these discrepancies can be attributed to assumptions of the model or its intrinsic limitations (see \citealt{2014ApJ...782...81V}). In this case, as discussed previously in Sect. \ref{sec_stars}, they arise mostly due to the spatial resolution of the surface field distributions. The overall lower densities of the corona and the imbalance of emission at different coronal temperatures, are directly related with the amount of confining loops and therefore, with the missing (un-resolved) surface magnetic field and its complexity. In addition, as will be presented in the Sect. \ref{sec_flux}, the simulated stellar X-ray and EUV luminosities appear underestimated. This may indicate that some adjustments are required in the coronal heating mechanism, when applying this particular model to resolution-limited surface field distributions (e.g. ZDI data). Further systematic work will be performed in this direction, analogous to the numerical grid presented in \cite{2014ApJ...783...55C}, including also other coronal emission ranges covered by current solar instrumentation (e.g. Solar Dynamics Observatory, \citealt{2012SoPh..275....3P}).

%The results from this quantitative analysis allow a more consistent calibration to the solar case (i.e. less dependant on the surface field resolution), as well for the comparison between the considered systems. 

\subsection{HD 1237 (GJ 3021)}\label{SC_HD1237}

\begin{figure*}[!ht]
\centering%  left, bottom, right and top
\includegraphics[scale=0.333]{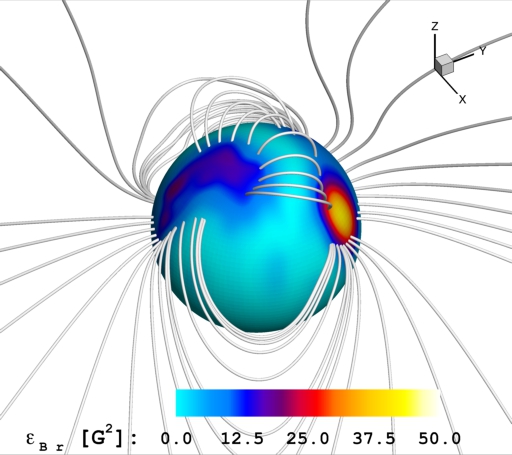}\hspace{2pt}\includegraphics[scale=0.333]{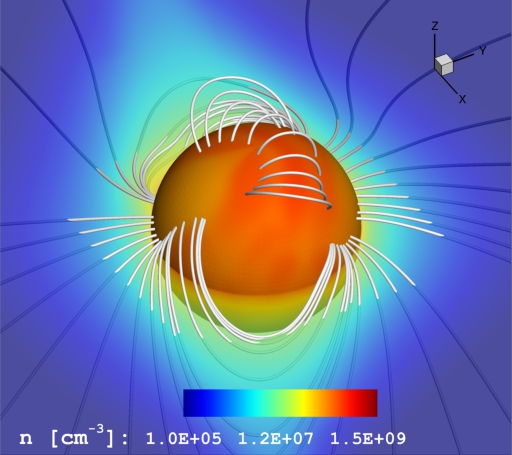}\hspace{2pt}\includegraphics[scale=0.333]{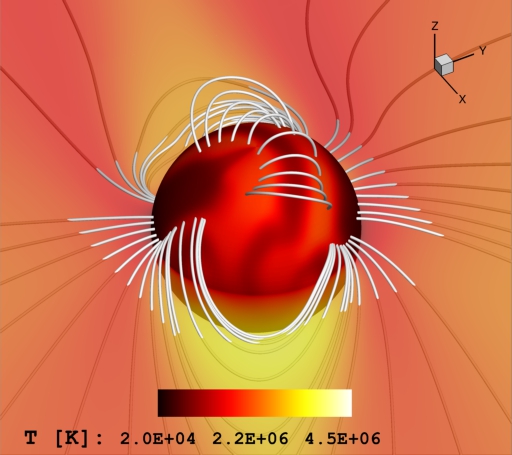}\vspace{1pt}
\includegraphics[scale=0.22505]{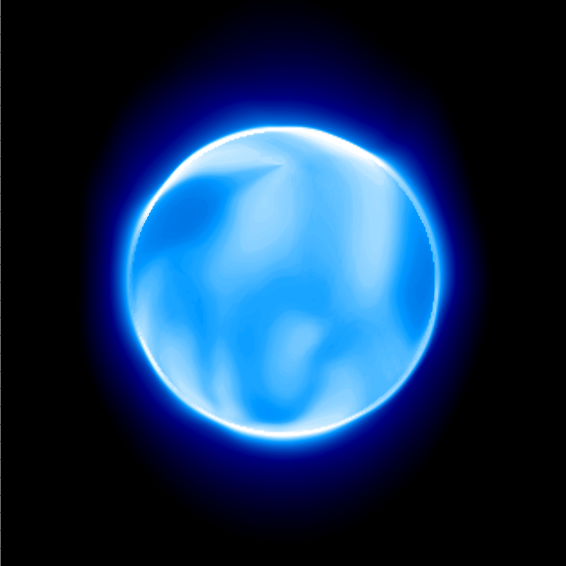}\hspace{2pt}\includegraphics[scale=0.22505]{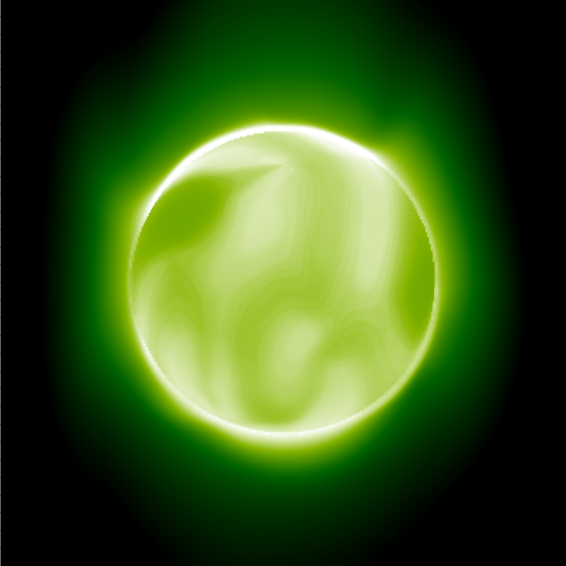}\hspace{2pt}\includegraphics[scale=0.22505]{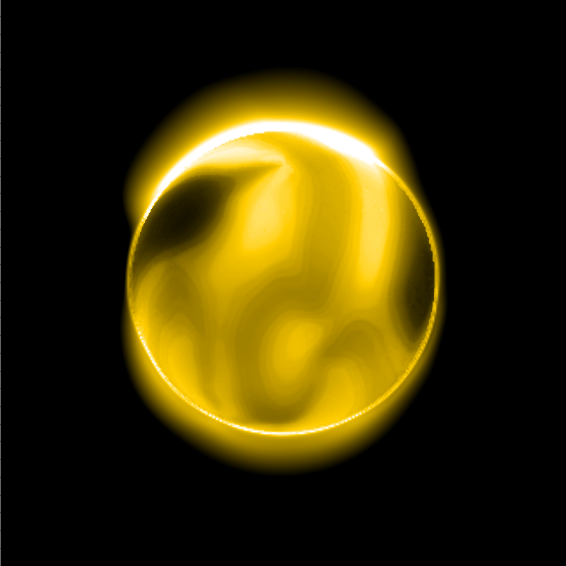}\hspace{2pt}\includegraphics[scale=0.22505]{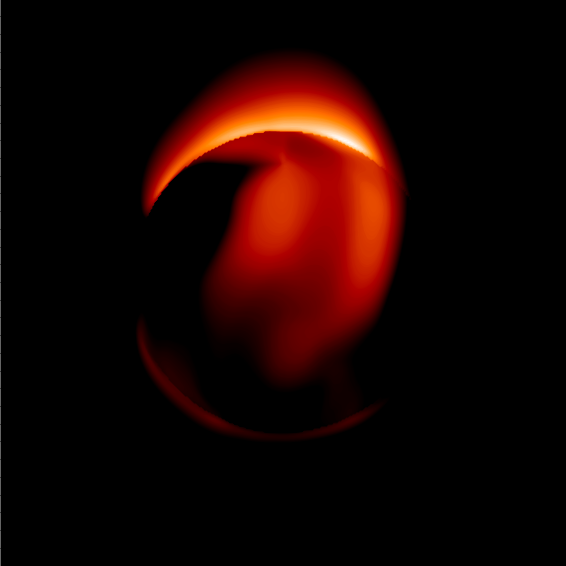}
\caption{Simulation results for the coronal structure of HD 1237 driven by the ZDI large-scale magnetic field map. The upper panels contain the distribution of the magnetic energy density ($\varepsilon_{\,B\,r}$, left), the number density ($n$, middle) and temperature ($T$, right). For the last two quantities the distribution over the equatorial plane ($z = 0$) is presented. The sphere represents the stellar surface and selected three-dimensional magnetic field lines are shown in white. The lower images correspond to synthetic coronal emission maps in EUV (blue / 171\,\AA\,, green / 195\,\AA\, and yellow / 284\,\AA) and SXR (red / 2 -- 30\,\AA). The perspective and colour scales are preserved in all panels, with an inclination angle of $i = 50\,^{\circ}$.}
\label{fig_5}
%\end{figure*}
\vspace{10pt}
%\begin{figure*}[!h]
\centering%  left, bottom, right and top
\includegraphics[scale=0.333]{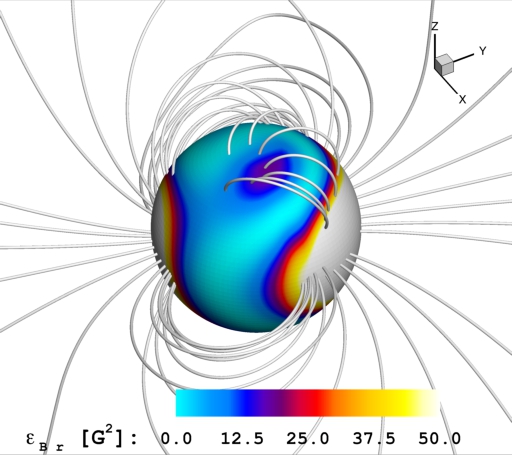}\hspace{2pt}\includegraphics[scale=0.333]{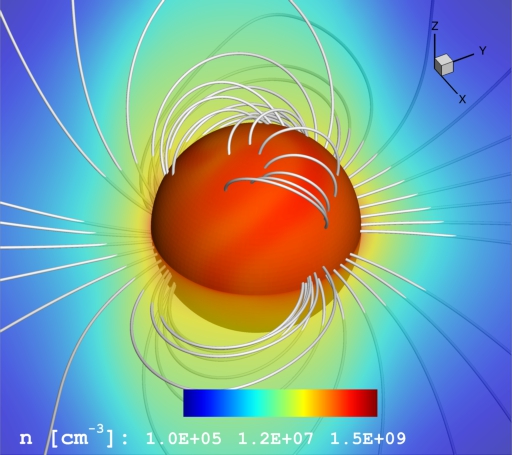}\hspace{2pt}\includegraphics[scale=0.333]{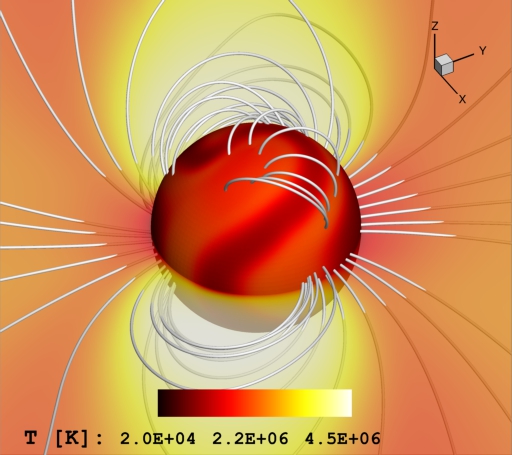}\vspace{1pt}
\includegraphics[scale=0.22505]{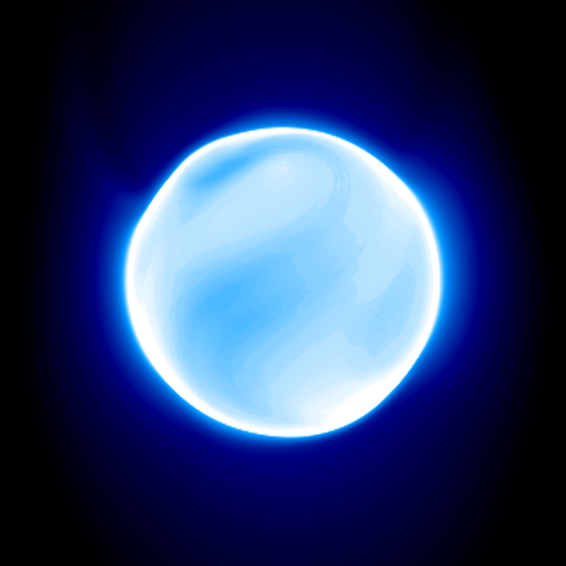}\hspace{2pt}\includegraphics[scale=0.22505]{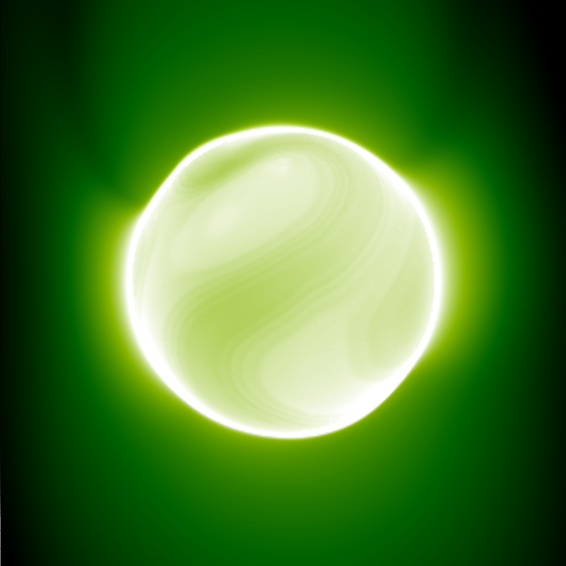}\hspace{2pt}\includegraphics[scale=0.22505]{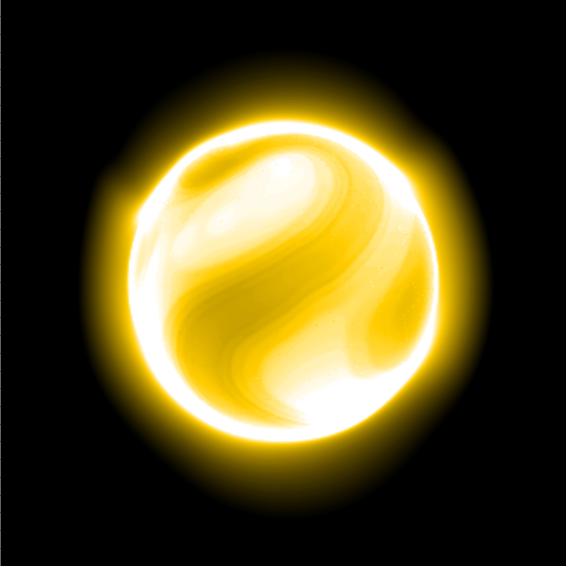}\hspace{2pt}\includegraphics[scale=0.22505]{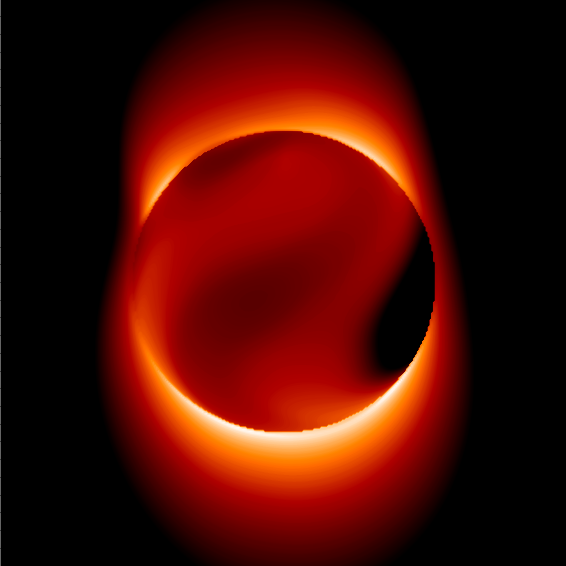}
\caption{Simulation results for the coronal structure of HD 1237 driven by the SH-ZDI large-scale field map. See caption of Fig. \ref{fig_5}. The three-dimensional magnetic field lines are calculated in the same spatial locations as in the solution presented in Fig. \ref{fig_5}.}
\label{fig_6}
\end{figure*}

The coronal structure obtained for HD 1237 shows a relatively simple topology. Two main magnetic energy concentrations, associated with the field distributions shown in Fig. \ref{fig_1}, dominate the physical properties and the spatial configuration in the final steady-state solution. The outer parts of these regions serve as foot-points for coronal loops of different length-scales. Close to the north pole an arcade is formed, which covers one of the main polarity inversion lines of the large-scale magnetic field. As can be seen in Figs. \ref{fig_5} and \ref{fig_6}, denser and colder material appears near these lines on the surface, resembling solar prominences or filaments. Larger loops extending higher in the corona, connect the opposite ends of both magnetic regions.

\clearpage

\begin{figure*}[!ht]
\centering%  left, bottom, right and top
\includegraphics[scale=0.333]{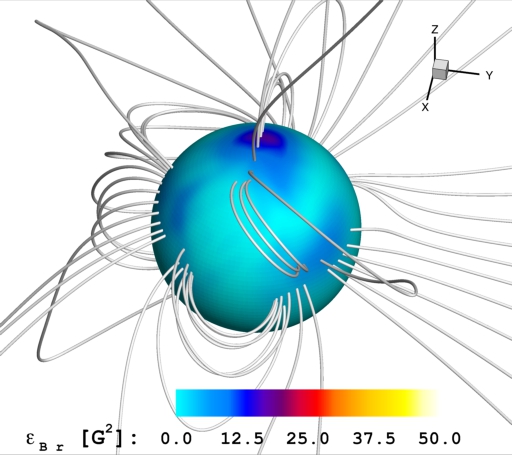}\hspace{2pt}\includegraphics[scale=0.333]{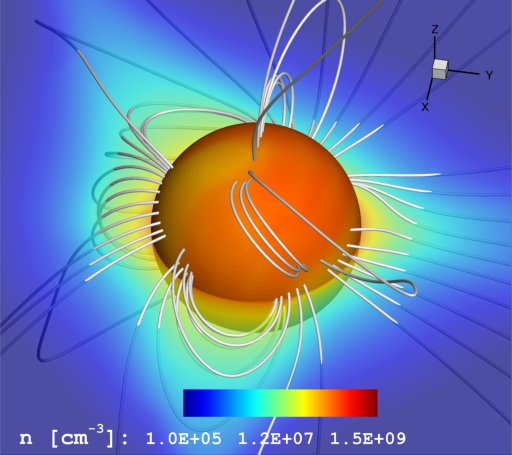}\hspace{2pt}\includegraphics[scale=0.333]{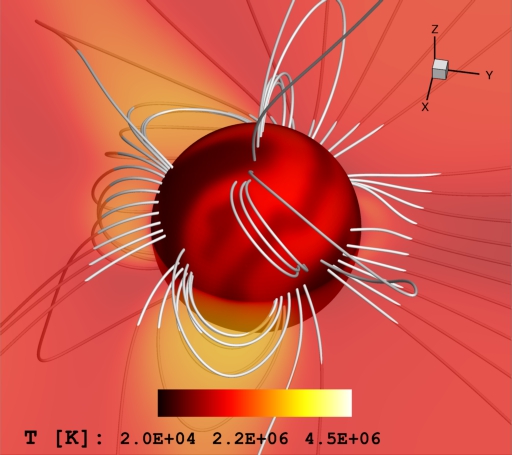}\vspace{1pt}
\includegraphics[scale=0.22505]{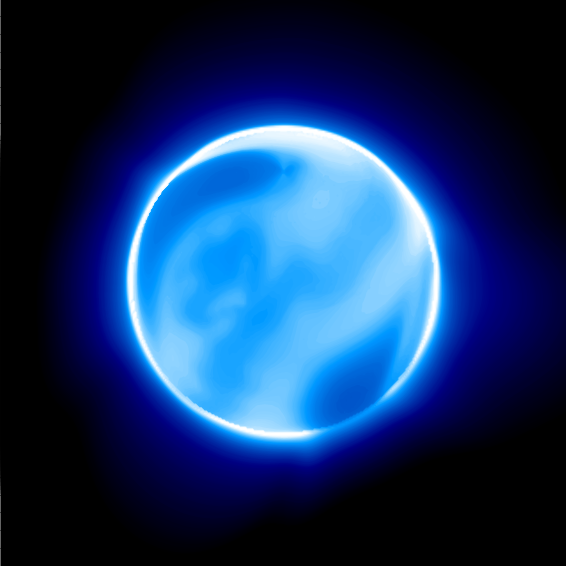}\hspace{2pt}\includegraphics[scale=0.22505]{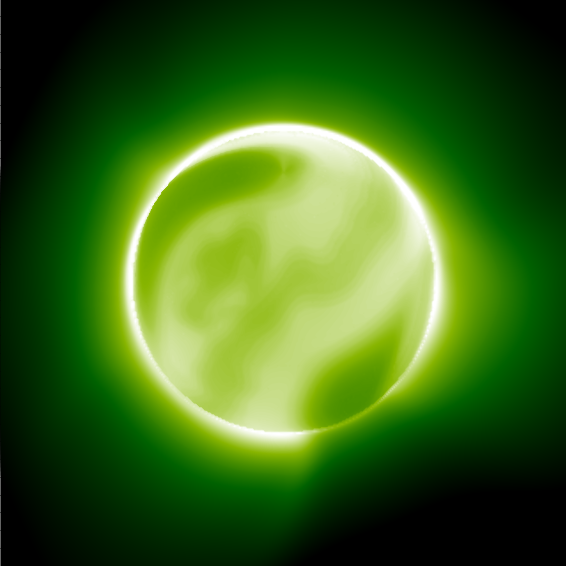}\hspace{2pt}\includegraphics[scale=0.22505]{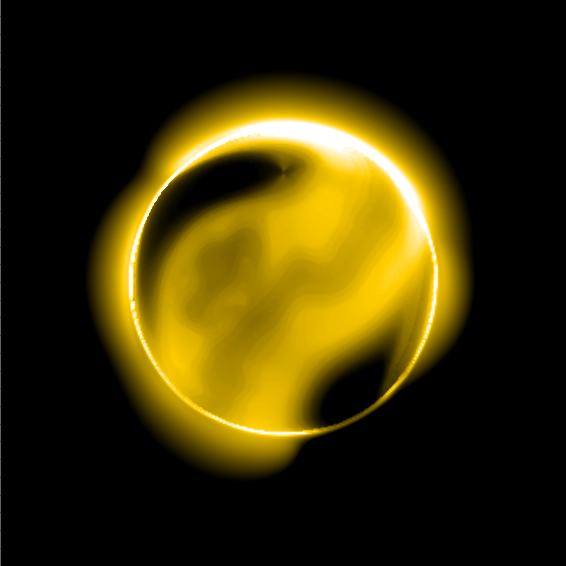}\hspace{2pt}\includegraphics[scale=0.22505]{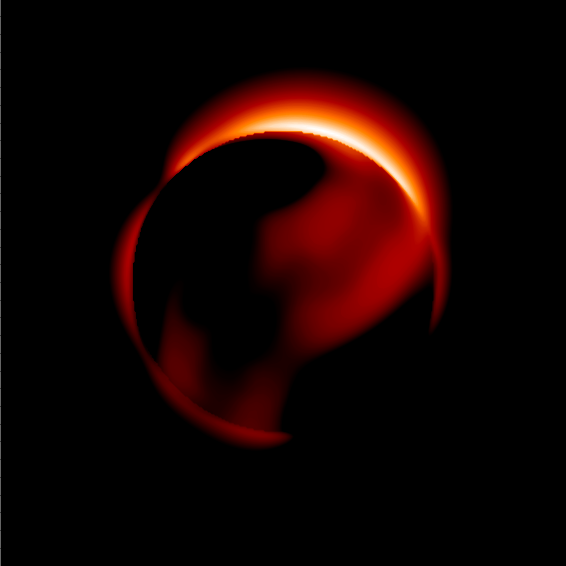}
\caption{Simulation results for the coronal structure of HD 22049 driven by the ZDI large-scale magnetic field map. The upper panels contain the distribution of the magnetic energy density ($\varepsilon_{\,B\,r}$, left), the number density ($n$, middle) and temperature ($T$, right). For the last two quantities the distribution over the equatorial plane ($z = 0$) is presented. The sphere represents the stellar surface and selected three-dimensional magnetic field lines are shown in white. The lower images correspond to synthetic coronal emission maps in EUV (blue / 171\,\AA\,, green / 195\,\AA\, and yellow / 284\,\AA) and SXR (red / 2 -- 30\,\AA). The perspective and colour scales are preserved in all panels, with an inclination angle of $i = 45\,^{\circ}$.}
\label{fig_7}
%\end{figure*}
\vspace{10pt}
%\begin{figure*}[!h]
\centering%  left, bottom, right and top
\includegraphics[scale=0.333]{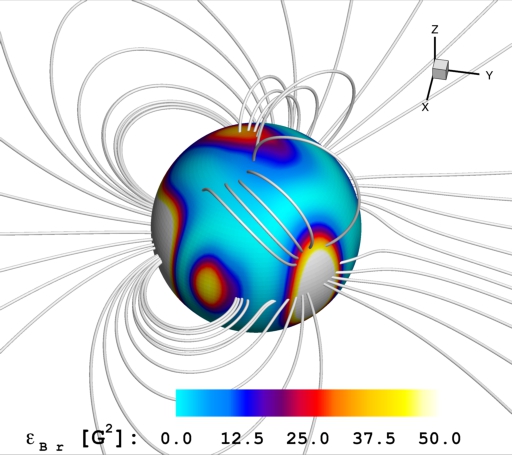}\hspace{2pt}\includegraphics[scale=0.333]{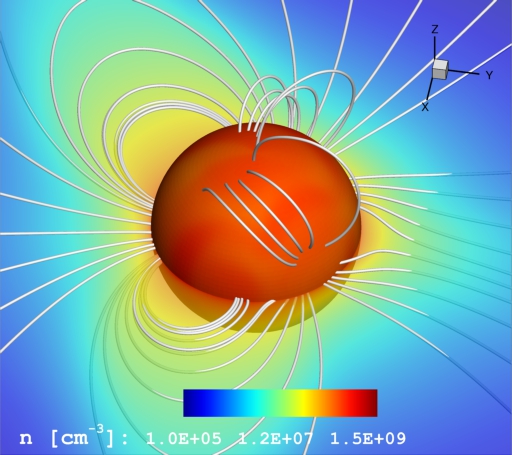}\hspace{2pt}\includegraphics[scale=0.333]{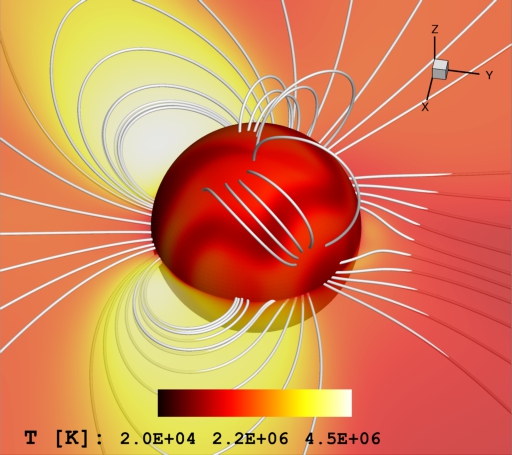}\vspace{1pt}
\includegraphics[scale=0.22505]{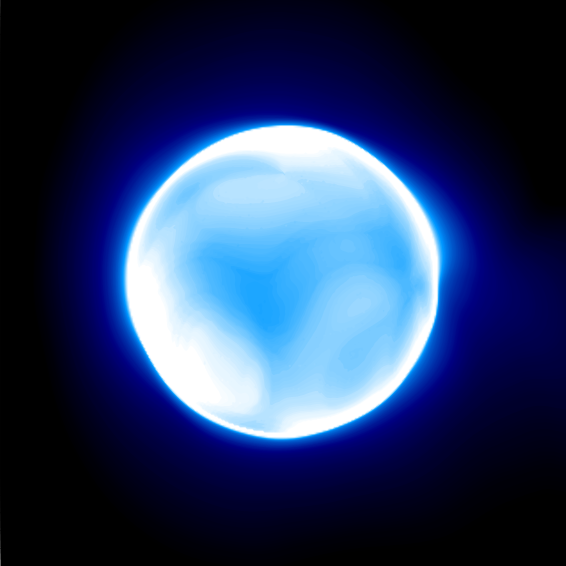}\hspace{2pt}\includegraphics[scale=0.22505]{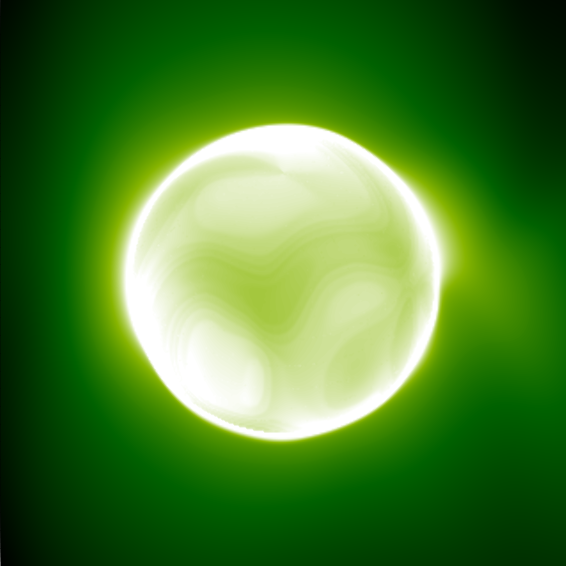}\hspace{2pt}\includegraphics[scale=0.22505]{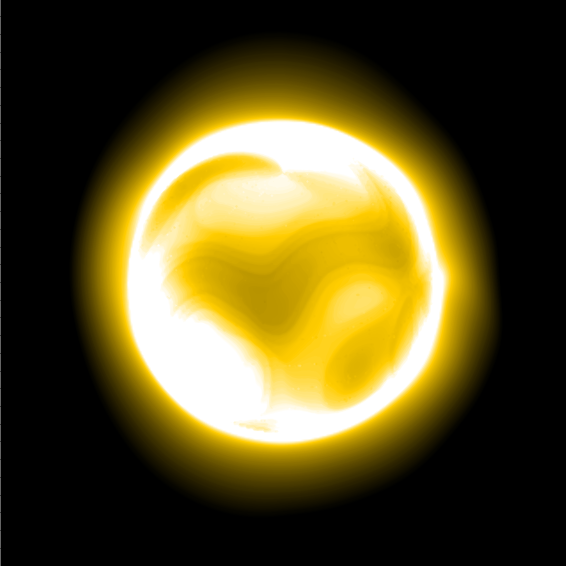}\hspace{2pt}\includegraphics[scale=0.22505]{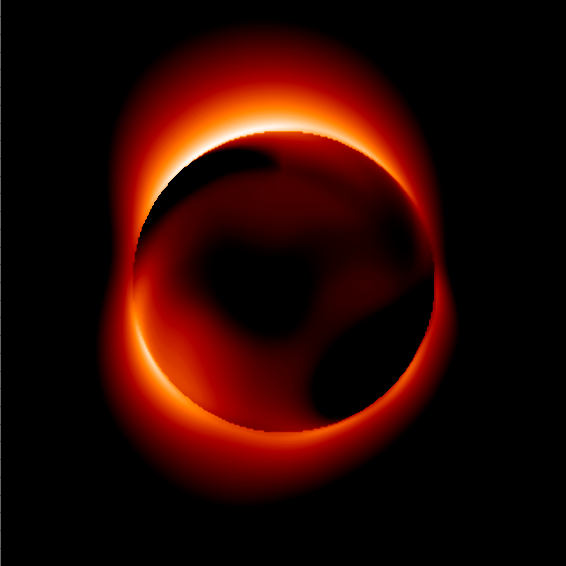}
\caption{Simulation results for the coronal structure of HD 22049 driven by the SH-ZDI large-scale field map. See caption of Fig. \ref{fig_7}. The three-dimensional magnetic field lines are calculated in the same spatial locations as in the solution presented in Fig. \ref{fig_7}}
\label{fig_8}
\end{figure*}

\clearpage

\begin{table*}[!htbp]
\caption{Average physical properties of the inner corona (IC) region (from 1.05 to 1.5 $R_*$).}             
\label{table_3}      
\centering
{\normalsize          
\begin{tabular}{ l | c c | c c | c | c c }    
\hline\hline
Parameter &  \multicolumn{2}{|c|}{HD 1237} & \multicolumn{2}{|c|}{HD 22049} & HD 147513 &  \multicolumn{2}{|c}{Sun}\\
 & ZDI & SH-ZDI & ZDI & SH-ZDI & SH-ZDI & CR 1922 (Min) & CR 1962 (Max) \\
\hline                    
$\left<n\right>_{\rm IC}$ [$\times\,10^{7}$ cm$^{-3}$] & 3.66 & 7.74 & 3.38 & 8.30 & 4.80 & 1.80 & 4.78\\
$\left<T\right>_{\rm IC}$ [$\times\,10^{6}$ K] & 2.49 & 3.42 & 2.06 & 3.20 & 2.79 & 1.48 & 2.07\\
$\left<B\right>_{\rm IC}$ [G] & 4.58 & 16.43 & 3.34 & 14.39 & 5.37 & 0.94 & 2.31\\
\hline                  
\end{tabular}}
\end{table*} 

\begin{figure*}[!ht]
\centering%  left, bottom, right and top
\includegraphics[scale=0.333]{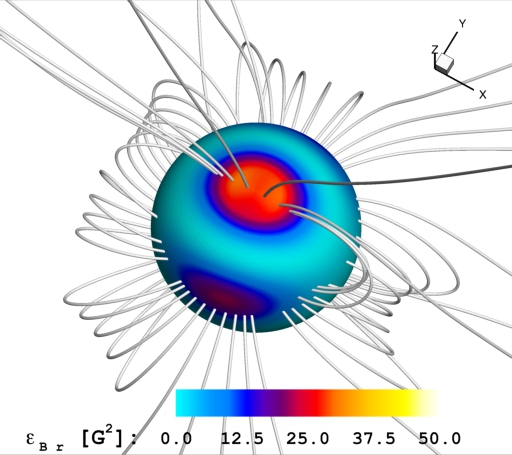}\hspace{2pt}\includegraphics[scale=0.333]{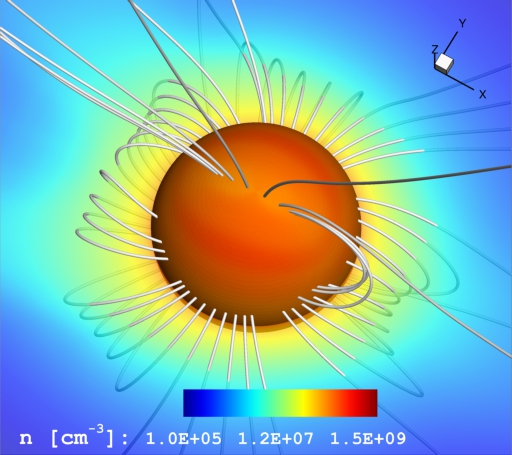}\hspace{2pt}\includegraphics[scale=0.333]{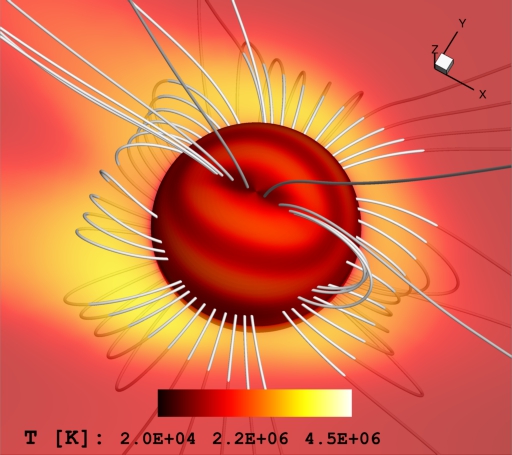}\vspace{1pt}
\includegraphics[scale=0.22505]{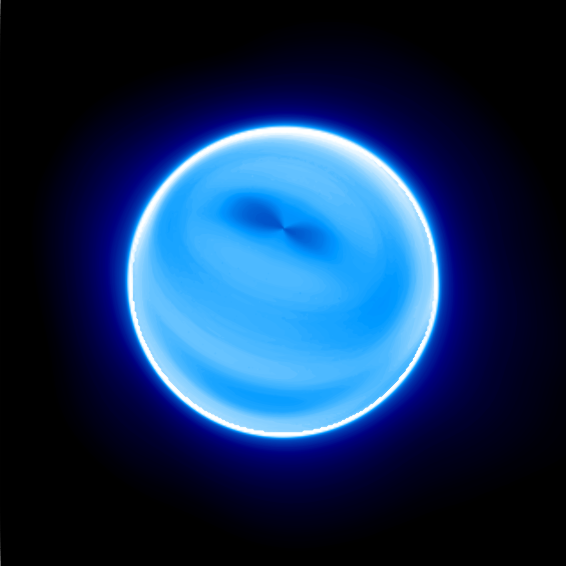}\hspace{2pt}\includegraphics[scale=0.22505]{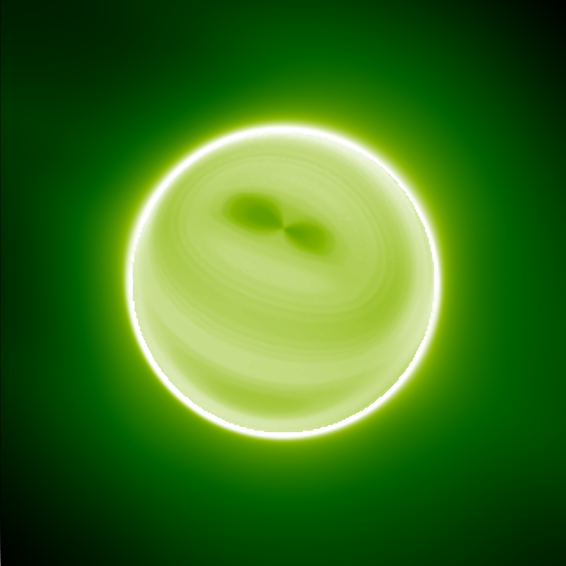}\hspace{2pt}\includegraphics[scale=0.22505]{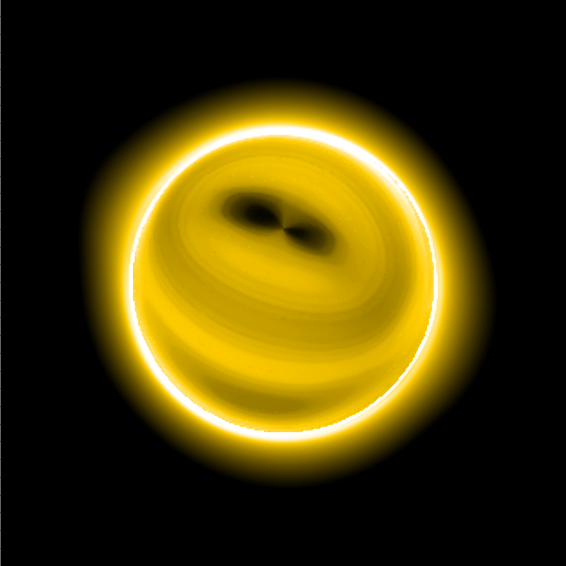}\hspace{2pt}\includegraphics[scale=0.22505]{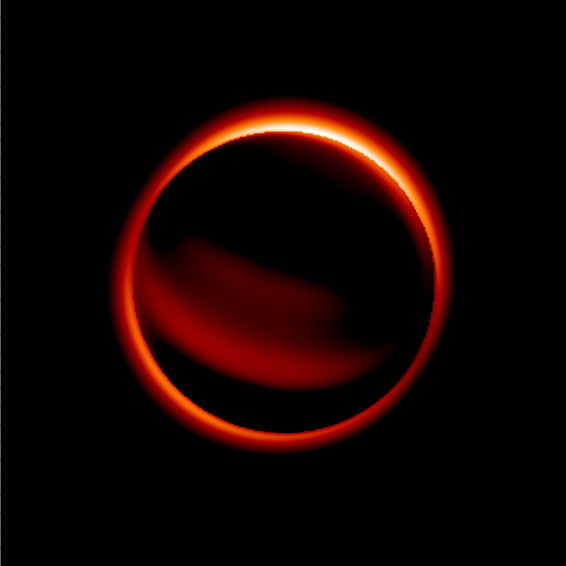}
\caption{Simulation results for the coronal structure of HD 147513 driven by the SH-ZDI large-scale magnetic field map. The upper panels contain the distribution of the magnetic energy density ($\varepsilon_{\,B\,r}$, left), the number density ($n$, middle) and temperature ($T$, right). For the last two quantities the distribution over the equatorial plane ($z = 0$) is presented. The sphere represents the stellar surface and selected three-dimensional magnetic field lines are shown in white. The lower images correspond to synthetic coronal emission maps in EUV (blue / 171\,\AA\,, green / 195\,\AA\, and yellow / 284\,\AA) and SXR (red / 2 -- 30\,\AA). The perspective is preserved in all panels, with an inclination angle of $i = 20\,^{\circ}$.}
\label{fig_9}
\end{figure*}

%\clearpage
\noindent These loops confine coronal material via magnetic mirroring, increasing the local density and temperature of the plasma. Some of this heated plasma is visible in the synthetic emission images of the lower corona (lower panels of Figs. \ref{fig_5} and \ref{fig_6}). 

Inside the two large magnetic energy regions, the coronal field lines are mainly open. This leads to the generation of coronal holes, where the material follows the field lines and leaves the star. In turn, this decreases the local plasma density and temperature in both regions, making them appear dark in the coronal emission maps. These coronal holes will have a strong influence in the structure of the stellar wind and the inner astrosphere. This will be discussed in detail in the second paper of this study. 

In terms of the field distribution (i.e. ZDI/SH-ZDI, Sect. \ref{sec_stars}), the global structure of the corona of HD 1237 is similar in both cases. This was expected since the largest features, in the surface field distributions, are common in both procedures. However, as can be seen directly in Figs. \ref{fig_5} and \ref{fig_6}, several qualitative and quantitative differences appear in various aspects of the resulting coronal structure. First, despite having the same thermodynamic base conditions, the SH-ZDI solution leads to a larger corona with an enhanced high-energy emission. This is a consequence of the available magnetic energy to heat the plasma, in combination with the size of the coronal loops (and therefore, the amount of material trapped by the field). 

To quantify these differences, we estimated the average density, temperature and magnitude of the coronal magnetic field, inside a spherical shell enclosing the region between 1.05 and 1.50 $R_*$. This range captures the bulk of the inner corona, with the lower limit selected to avoid possible numerical errors in the average integration (due to the proximity with the boundary of the simulation domain). The integrated values obtained for each parameter, and for the other stars, are listed in the Table \ref{table_3}. %This procedure allows a broad comparison between the numerical solutions of the considered stars (and with the solar case). 
%Table from next section.

For HD 1237 we obtain differences by a factor of $\sim$1.4 in temperature, $\sim$2.1 in density and $\sim$3.5 in magnetic field strength, among both cases. As the corona is hotter and denser in the SH-ZDI case, the resulting high-energy emission is almost featureless in the EUV channels ($T \sim 1 - 2$ MK). In addition, the impact of the surface field completeness is clear in the SXR image, where the coronal holes are shifted to lower latitudes and the emission comes from both hemispheres of the star (in contrast to the simulated emission in this range for the ZDI case). 

As expected, HD 1237 shows enhanced coronal conditions compared to the Sun, especially for the SH-ZDI case (see Table \ref{table_3}). For the ZDI case the mean coronal density appears to be lower than the Solar maximum value (by $\sim$25\%). This may be connected with the incompleteness of the ZDI maps (Sect. \ref{sec_stars}), since a similar situation occurs for the ZDI solution of HD 22049 by roughly the same amount.

\subsection{HD 22049 ($\epsilon$ Eridani)}\label{SC_HD22049}

\noindent The solutions for HD 22049 are presented in Figs. \ref{fig_7} and \ref{fig_8}. The coronal structure in this case is highly complex, with several hot and dense loops connecting the different polarity regions of the surface field distribution. In some locations, the material is able to escape near the cusp of the loops, resembling helmet streamers in the Sun. For the SH-ZDI simulation, some of this escaping material is even visible in the EUV synthetic maps (in particular in the $195\,\AA$ channel -- Green image in Fig. \ref{fig_8}). 

Similar to HD 1237, two large coronal holes are visible in the synthetic high-energy emission maps (especially in the ZDI simulation). However in this case, the correlation with the stronger magnetic features in the surface is less clear as for HD 1237. A large filament crossing the entire disk is visible in both solutions, being more smooth in the SH-ZDI as is expected from the underlying field distribution. 

The comparison between the ZDI and SH-ZDI solution leads to similar results as for HD 1237. The variation in the average coronal density, temperature and magnetic field strength reach factors of 2.5, 1.6 and 4.3, respectively (see Table \ref{table_3}). The differences in the synthetic emission maps are also somewhat preserved with respect to the HD 1237 simulations; Less coronal features are visible in EUV channels of the SH-ZDI solution, and the SXR emission is dominated by the closed field regions, distributed in this case in various locations of the three-dimensional structure.    

Finally, it is interesting to note here the similarities between the quantitative average properties of the ZDI solution of HD 22049 and the solar maximum case. The resulting mean temperatures and field strengths are commensurate among these simulations. However, large differences are evident in the qualitative aspects of both solutions (see Figs. \ref{fig_7} and \ref{fig_A2}). No coronal holes are obtained for the solar maximum case, and the high-energy emission is highly concentrated from small portions of the corona (associated with active regions). This again can be understood in terms of the amount of magnetic structures resolved in the surface field distribution. Despite the degraded resolution for the solar case, the number of bipolar regions on the surface (sustaining dense coronal loops) is much larger than in the large-scale field maps recovered with ZDI. Instead, the ZDI coronal solution for HD 22049 is much more similar to the solar minimum case (Fig. \ref{fig_A1}). This clearly exemplifies the importance of combining quantitative descriptions, together with qualitative spatially-resolved information for a robust comparison.  

\subsection{HD 147513 (HR 6094)}\label{SC_HD147513}

\noindent We present the steady-state coronal solution for HD 147513 in Fig. \ref{fig_9}. As was mentioned earlier, we only consider the SH-ZDI field distribution in this case (see Sect. \ref{sec_stars}). The coronal structure is dominated by a rather simple configuration of poloidal loops, driven by the surface field distribution (mainly from the dipolar and quadrupolar components). This generates bands of trapped material, separated by the magnetic polarity inversion lines and distributed at different latitudes. Few open field regions are visible in the coronal structure, which are again located inside the largest magnetic energy concentrations. One of these regions appears in the north pole of the star, which suffers a small distortion in the EUV images due to a numerical artifact of the spherical grid. The line-of-sight SXR emission displays a ring-like structure close to the limb, corresponding to the hottest material of the steady-state corona. Some faint emission can be also seen inside the stellar disk. As the estimated inclination angle for this star is small ($i\sim$20$^{\circ}$), the coronal features are visible at almost all rotational phases.        

The coronal properties listed in Table \ref{table_3}, show an average density comparable to the solar case in activity maximum. However, as was presented in Sect. \ref{SC_Calibration}, the limited resolution of the surface field distribution can strongly affect this parameter. Given the relatively low resolution for the SH-ZDI map for this star, we expect larger discrepancies than the ones obtained for the solar case. In this sense, the average values obtained from the simulation correspond only to rough estimates of the actual conditions of the corona. This is considered in more detail in Sect. \ref{cor_prop}. Still, the geometrical configuration of this system provides an interesting view of the coronal features, that cannot be easily obtained even for the solar case.

\section{Analysis and Discussion}\label{sec_discussion}

\noindent Using the simulation results we can relate the characteristics of the surface field distributions, with the obtained coronal properties and the environment around these systems. We will focus our discussion on three main aspects, including the thermodynamic structure, the coronal high-energy emission, and the stellar rotational modulation of the coronal emission.
 
\subsection{Thermodynamic Coronal Properties}\label{cor_prop}

\begin{figure}[t]
\centering %  left, bottom, right and top
\includegraphics[trim=0.0cm 0.0cm 0.0cm 0.0cm, clip=false, width=\hsize]{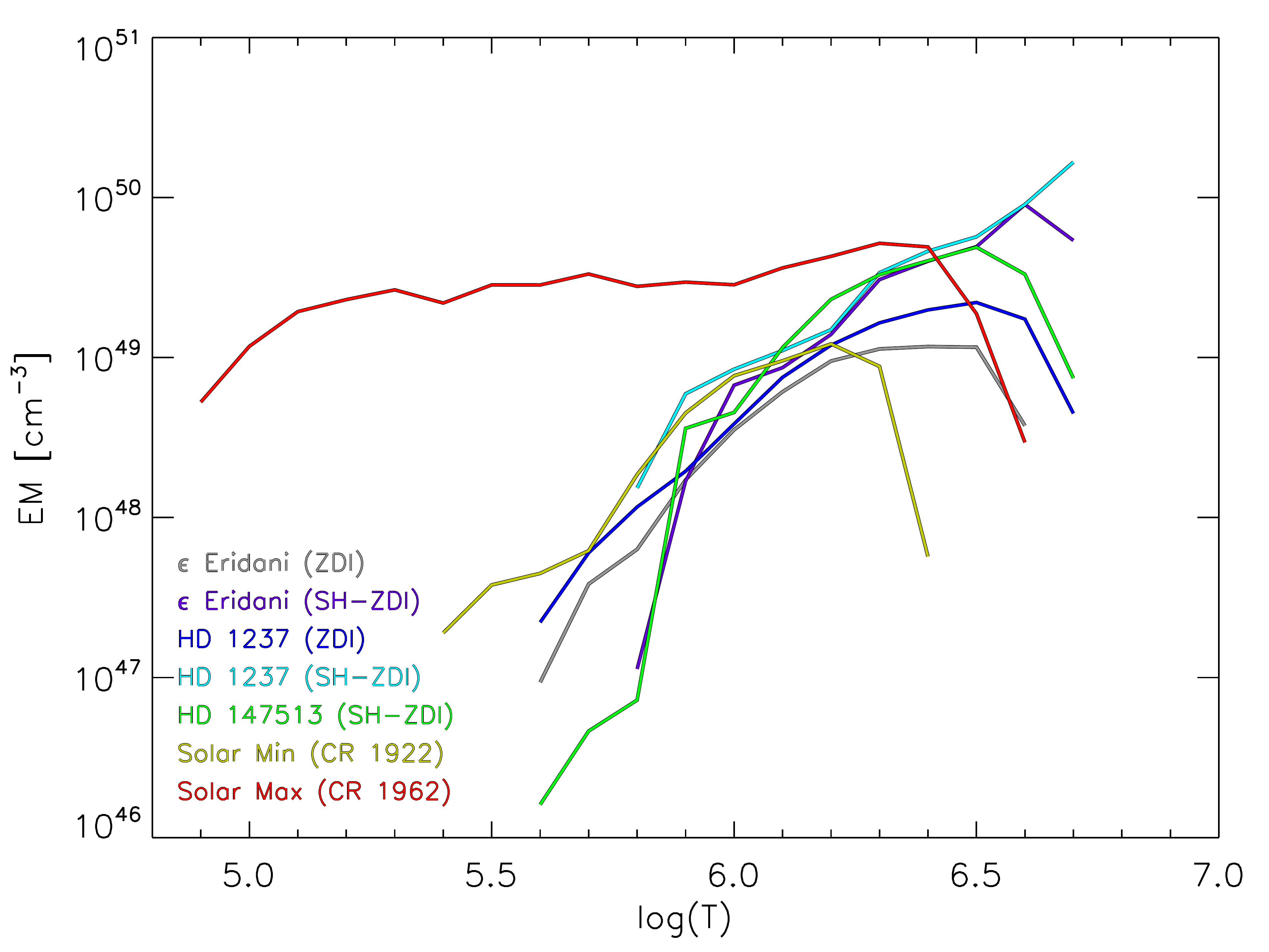}
%\caption{Simulated inner coronal temperature ($T$) and emission measure ($EM$) as a function of the unsigned radial magnetic flux $\left<\Phi_{\rm Br}\right>_{\rm s}$. Note the log-scale in the right-axis. See the text for more details.}
\caption{Emission measure distributions $EM(T)$ calculated from the 3D steady-state solutions. Each colour corresponds to one of the simulations presented in Sect.~\ref{sec_results}, including the solar runs (Appendix \ref{app_1}).}
\label{fig_10}
\vspace{-14pt}
\end{figure}

\noindent From the simulated 3D structure in each star, we calculate the emission measure distribution, $EM(T)$, defined by

\begin{equation}
 EM(T) = \int_{V(T)} n^2(T)dV(T)\text{ ,} 
\end{equation}

\noindent where $n(T)$ is the plasma density at the temperature $T$, the integration only includes the volume of the grid cells at that particular temperature, and the volume covers all the closed field line regions in the steady-state solutions. We use temperature bins of 0.1 in $\log T$ starting from the base temperature (i.e. $\log T \simeq 4.9$), up to the maximum temperature achieved in each simulation. Figure \ref{fig_10} contains the computed $EM(T)$ for all the considered cases. As expected, the peak values are located at $\log T > 6.0$, and move towards larger emission measures and higher temperatures, with increasing average (radial) magnetic energy density $\left<\varepsilon_{B\,r}\right>$ (see Table \ref{table_1}).

%\noindent Finally, we investigated how the thermodynamic properties of the corona change with the amount of large-scale surface magnetic flux. Figure \ref{fig_12} contains the relations obtained for the coronal temperature ($T$) and the emission measure ($EM$), in the inner corona region. These values correspond to the three-dimensional averages discussed previously in Sect. \ref{sec_results} (see also Table \ref{table_3}).

%As indicated by the continuous lines in Fig. \ref{fig_12}, both quantities seem to scale following a log-relation with a power of $\left<\Phi_{\rm Br}\right>_{\rm s}$, albeit with considerable scatter for the average emission measure $\left<EM\right>_{\rm IC}$. Despite the differences in their calculation, the power-law relations are consistent with the results obtained for the high-energy coronal emission (Sect. \ref{sec_flux}). As expected, the temperature distribution follows the SXR emission (hotter component), while the emission measure appears more sensitive to the behaviour in the EUV range (denser component).

In a similar manner to the solar case (Sect.~\ref{SC_Calibration}), we compare the simulated quantities to observational values. The ZDI and SH-ZDI simulations of HD~22049 yield maximum $EM$ values of $\log EM$\,$\simeq$\,49.1 (at $\log T$\,$\simeq$\,6.4) and $\log EM$\,$\simeq$\,50.0 (at $\log T$\,$\simeq$\,6.6), respectively. The peak temperature and emission measure of the ZDI model are significantly lower than those derived from both EUV and X-ray spectra ($\log EM$\,$\simeq$\,50.7 at $\log T$\,$\simeq$\,6.6\,$\pm$\,0.05, \citealt{2000ApJ...545.1074D,2004A&A...416..281S,2008MNRAS.385.1691N}). The SH-ZDI emission measure fares somewhat better, with good agreement in terms of the peak temperature. However, this model still predicts an emission measure significantly lower than observed, by roughly a factor of 5.

For HD 147513 available observations, from the broad-band filters of the {\it Extreme Ultraviolet Explorer} (EUVE) Deep Survey telescope, only provide rough estimates of the coronal conditions, suggesting a probable emission measure in the range $\log EM\sim 51$--52  \citep{1993ApJ...414L..61V} but with no discrimination on the temperature.
%with an average value of $\log(EM)$\,$\sim$\,52 at $\log(T)$\,$\sim$\,6.4 . 
In turn, the peak of the simulated distribution is located at $\log T$\,$\simeq$\,6.5, with an associated value of $\log EM$\,$\simeq$\,49.7.  The discrepancy in emission measure  might be expected given the relatively low spatial resolution of the SH-ZDI map driving the simulation (Sections \ref{sec_stars} and \ref{SC_HD147513}). The peak temperature is also slightly lower than what might be expected based on the emission measure distribution and the observed peak temperature of HD~22049.

In the case of HD~1237, there are no observational constraints in the literature regarding the $EM$ distribution. From the numerical simulations, we obtain peak values of $\log EM$\,$\simeq$\,49.3 at $\log T$\,$\sim$\,6.5 for the ZDI case, and $\log EM$\,$\simeq$\,50.2 at $\log T$\,$\sim$\,6.7 for the SH-ZDI case. 
 
In all the stellar cases, the simulated $EM$ distributions show maxima close to the expected values for stars within the considered levels of activity (see Table \ref{table_1}). However, the emission measures are systematically lower than indicated by observation.

%However, as the simulations were performed consistently for all the considered cases (see sections \ref{sec_stars} and \ref{sec_num}), we can still trace the overall behaviour of the coronal properties from the numerical results. 

The behaviour of the simulated $EM$ distribution for the solar maximum case (red line in Fig. \ref{fig_10}) is particularly interesting, compared with the remaining simulations. Both the peak emission measure and temperature are in good agreement with assessments from full solar disk observations \citep[e.g.][]{1995ApJ...443..416L,2000ApJ...545.1074D}. However, the observations indicate a slope in the $EM$ vs. temperature of order unity or greater, whereas the model prediction is much flatter. This results in a substantial over-prediction of the cooler emission measure at temperatures $\log T \le 6$ compared with observations. The solar minimum $EM$ distribution (yellow line in Fig. \ref{fig_10}) is more similar to the stellar cases in this regard. These differences have a considerable impact in the predicted coronal emission, as discussed in the next section.

\subsection{High-Energy Emission and Magnetic Flux}\label{sec_flux}

\begin{figure}[ht]
\centering %  left, bottom, right and top
\includegraphics[trim=0.0cm 0.0cm 0.0cm 0.0cm, clip=false, width=\hsize]{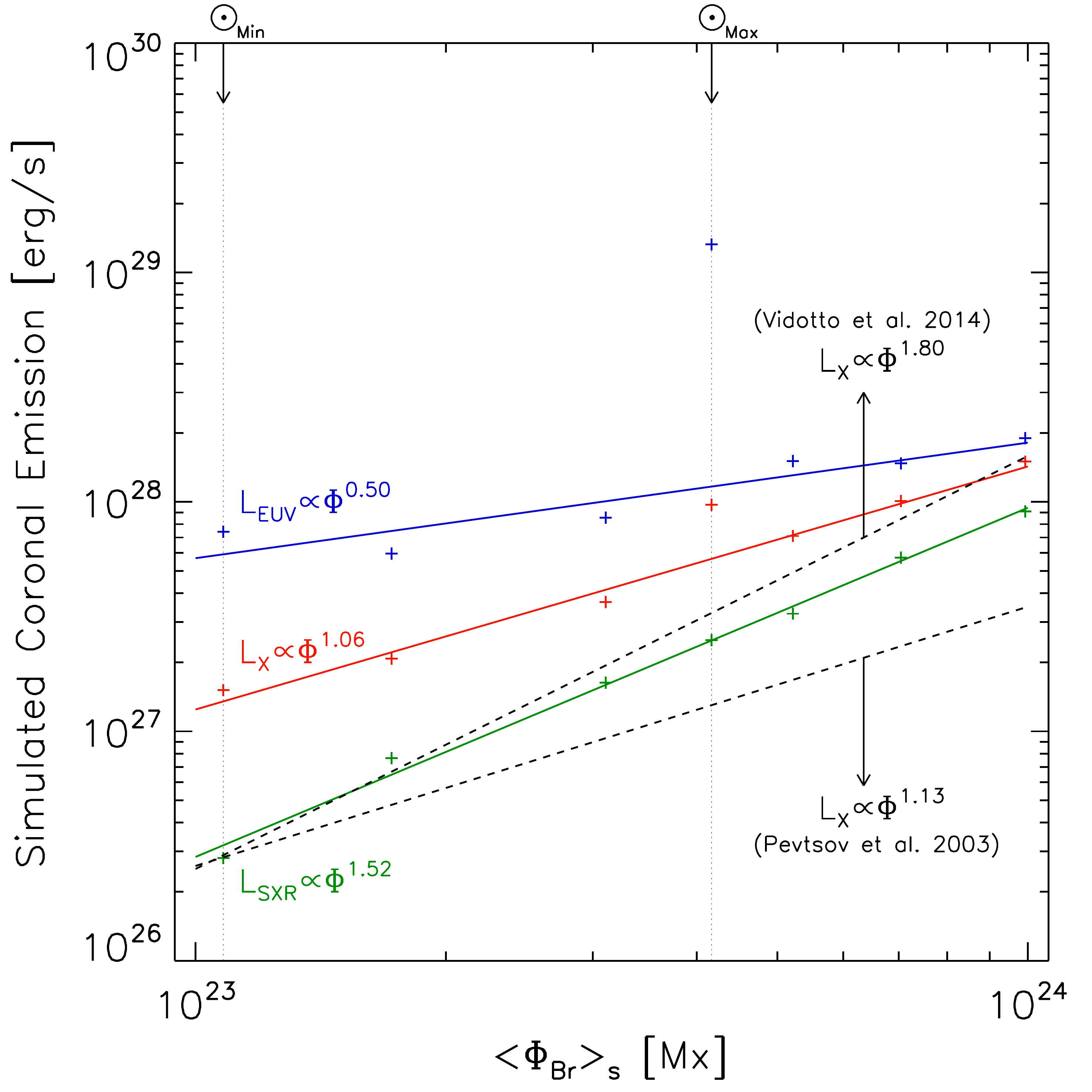}
\caption{Simulated high-energy coronal emission vs. unsigned radial magnetic flux $\left<\Phi_{\rm Br}\right>_{\rm s}$. Each point corresponds to one of the simulations described in Sect.~\ref{sec_results}, including the solar cases as indicated. These values are calculated from synthetic spectra, based on the $EM(T)$ distributions (Sect. \ref{cor_prop}), and integrated in the SXR ($2 - 30\,\AA$, green), X-ray ($5 - 100\,\AA$, red) and EUV ($100 - 920\,\AA$, blue) bands. The solid lines correspond to fits to the simulated data points. The dashed lines are based on observational studies using X-ray, against magnetic field measurements using ZB \citep{2003ApJ...598.1387P} and ZDI \citep{2014MNRAS.441.2361V}.}
\label{fig_11}
\vspace{-8pt}
\end{figure}

An observational study performed by \citet{2003ApJ...598.1387P} showed a relation between the unsigned magnetic field flux, $\Phi_{\rm B}$, and the X-ray emission, $L_{\rm X}$, covering several orders of magnitude in both quantities ($L_{\rm X} \propto \Phi^{1.13 \pm 0.05}_{\rm B}$). The analysis included various magnetic features of the Sun, together with Zeeman Broadening (ZB) measurements of active dwarfs (spectral types F, G and K), and pre-main sequence stars (see \citealt{1996IAUS..176..237S}). More recently, \citet{2014MNRAS.441.2361V} investigated the behaviour of various astrophysical quantities, including $L_{\rm X}$, with respect to the large-scale magnetic field flux (recovered with ZDI). They also found a power-law relation for both parameters ($L_{\rm X} \propto \Phi^{1.80 \pm 0.20}_{\rm B}$). These observational results have been interpreted as an indication of a similar coronal heating mechanism among these types of stars. 

%In this context, we have considered this relation from a numerical point of view, by using our simulated high-energy emission maps (e.g. EUV and SXR) and the underlying surface magnetic field flux distributions ($\Phi_{\rm B} = 4\pi |B_{r}| R^{2}_{*}$). In this analysis, we have included the results from all the considered cases (e.g. solar and ZDI/SH-ZDI), treating the solutions independently. This allows us to explore a broader range for both parameters, while maintaining the considerations and limitations of the data-driven numerical approach. In principle, this can be also studied from a more generic numerical point of view (i.e. including different simulated field distributions). However, this would require implicit assumptions about the field strength and spatial configuration (mostly influenced by the solar case), introducing strong biases in the analysis. Therefore considering the different recovered field maps (e.g. ZDI/SH-ZDI) as independent observations, represents a reasonable approximation. 

In this context, we have considered this relation from a numerical point of view, by simulating the coronal high-energy emission (based on the $EM(T)$ distributions presented in the previous section) and comparing the predicted fluxes with the underlying surface magnetic field flux distributions ($\Phi_{\rm B} = 4\pi |B_{r}| R^{2}_{*}$). In this analysis, we have included the results from all the considered cases (e.g. solar and ZDI/SH-ZDI), treating the solutions independently. This allows us to explore a broad range for both parameters, while maintaining the considerations and limitations of the data-driven numerical approach. In principle, this can be also studied from a more generic numerical point of view (i.e. including different simulated field distributions). However, this would require implicit assumptions about the field strength and spatial configuration (mostly influenced by the solar case), introducing strong biases in the analysis. Therefore considering the different recovered field maps (e.g. ZDI/SH-ZDI) as independent observations, represents a reasonable approximation. 

\begin{figure*}[!ht]
\centering%  left, bottom, right and top
\includegraphics[trim=0.0cm 0.0cm 4.0cm 0.5cm, clip=true, scale=0.3]{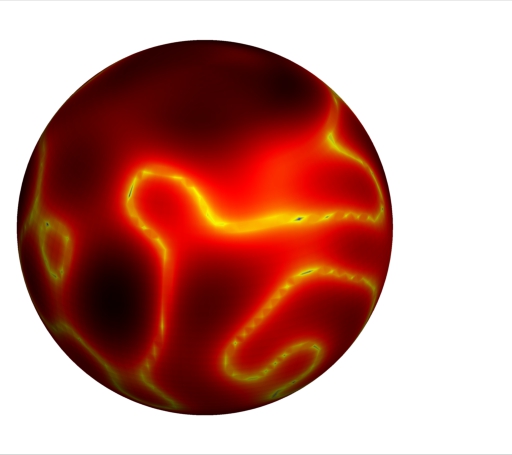}\hspace{0pt}\includegraphics[trim=0.0cm 0.0cm 4.0cm 0.5cm, clip=true, scale=0.3]{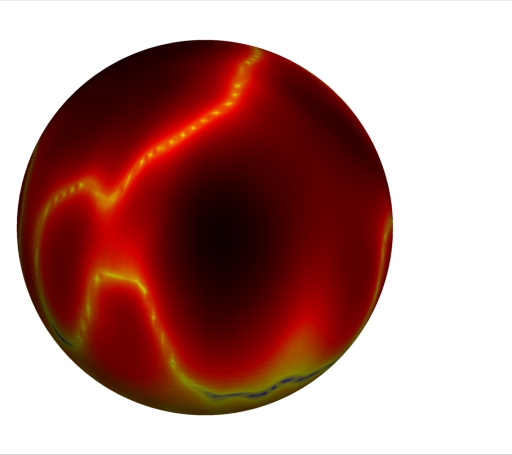}\hspace{0pt}\includegraphics[trim=0.0cm 0.0cm 4.0cm 0.5cm, clip=true, scale=0.3]{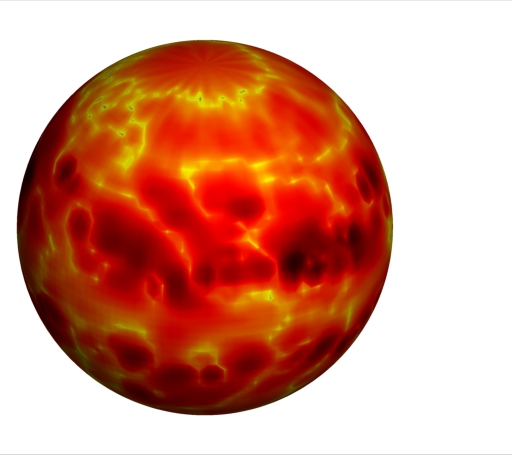}\hspace{1pt}
\includegraphics[trim=0.5cm 0.0cm 0.0cm 0.5cm, clip=true, scale=0.3]{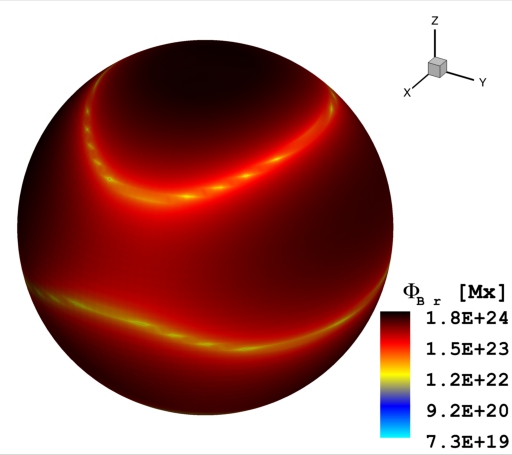}
\caption{Spatial distribution of the surface radial magnetic flux $\Phi_{\rm Br}$. The average surface magnetic flux increases from left to right, showing the Sun (Min), HD 1237 (ZDI), Sun (Max), and HD 147513 (SH-ZDI). The degree of complexity is clearly higher in the solar maximum case, despite the degraded spatial resolution. The perspective is preserved in all cases with an inclination angle of $i = 60\,^{\circ}$.}
\label{fig_12}
\vspace{0pt}
\end{figure*}

Spectra were simulated for each of the emission measure distributions, $EM(T)$, over the X-ray and EUV wavelength regimes, from 1 to 1100~\AA\ on a 0.1~\AA\ grid, covering all the bandpasses of interest to this work. Emissivities were computed using atomic data from the CHIANTI database version 7.1.4 \citep{1997A&AS..125..149D,2013ApJ...763...86L} as implemented in the Package for INTeractive Analysis of Line Emission (PINTofALE)\footnote[2]{\url{http://hea-www.harvard.edu/~PINTofALE/}}. The radiative loss in the temperature range of interest for the stars in this study is dominated by metals---principally Si, Mg and Fe---and the chemical abundance mixture has a concomitant influence on the predicted EUV and X-ray fluxes. HD~22049 and other intermediate activity G and K dwarfs exhibit a solar-like ``first ionization potential effect'' in which elements with low first ionization potentials ($\leq 13.6$~eV) can be enhanced by factors of up to 4 in the corona relative to photospheric values \citep[e.g.][]{1996ApJ...462..948L,2010ApJ...717.1279W}.  For the purposes of this study we do not try to match these abundances but instead adopt the solar abundance mixture of \citet{1998SSRv...85..161G} as a standard reference set.  Fluxes in the different bandpasses discussed below were obtained by integrating the synthetic spectra within the wavelength limits of interest. 

%\noindent Figure \ref{fig_11} contains the relation between the simulated high-energy coronal emission and the unsigned magnetic flux from the radial magnetic field maps. The plotted values correspond to a particular average for each parameter. The unsigned radial magnetic flux, noted as $\left<\Phi_{\rm Br}\right>_{\rm s}$, has been averaged over the entire surface of the star. For the high-energy emission, we consider an average over rotation, keeping the maximum and minimum variations as error bars. Two vertical arrows in the upper x-axis, denote the results for the solar case ($\odot_{\rm Min}$ and $\odot_{\rm Max}$). The different colours correspond to each of the synthetic emission maps presented in Sect. \ref{sec_results}. As indicated, a power-law fit has been applied for each channel (continuous lines). The segmented lines correspond to the previous observational results from \citet{2003ApJ...598.1387P} and \citet{2014MNRAS.441.2361V}. We have preserved an absolute scale in the y-axis, so no additional shifts have been included between any of the fits.  

Figure \ref{fig_11} shows the relation between the simulated high-energy coronal emission and the unsigned magnetic flux from the radial magnetic field maps. The latter, noted as $\left<\Phi_{\rm Br}\right>_{\rm s}$, has been averaged over the entire surface of the star.  The colours correspond to the spectral ranges used for the integration,  covering the SXR ($2 - 30\,\AA$, green), X-ray ($5 - 100\,\AA$, red) and EUV ($100 - 920\,\AA$, blue) bands. As indicated, a power-law fit has been applied to each channel (continuous lines), while the segmented lines correspond to the previous observational results from \citet{2003ApJ...598.1387P} and \citet{2014MNRAS.441.2361V}\footnote[3]{We have shifted the relation from \citet{2014MNRAS.441.2361V} to match the \citet{2003ApJ...598.1387P} relation at the Solar minimum value. This is performed for comparison purposes, as there are still discrepancies in the absolute values of the observational $L_{\rm X} - \Phi_{\rm B}$ relations, most likely connected with the method to estimate the surface magnetic flux (i.e. ZB in \citealt{2003ApJ...598.1387P} and ZDI in \citealt{2014MNRAS.441.2361V}).}. Two vertical arrows in the upper x-axis, denote the results for the solar case ($\odot_{\rm Min}$ and $\odot_{\rm Max}$).

Several aspects of Fig.~\ref{fig_11} are noteworthy. First, despite the reduced range in $\left<\Phi_{\rm Br}\right>_{\rm s}$, we were able to retrieve a similar behaviour between radiative flux and magnetic flux as in previous observational studies with larger datasets. For the simulated X-ray range, the results are consistent with the relation obtained by \citet{2003ApJ...598.1387P}, while in the SXR range the power-law dependance is more similar to the results obtained by \citet{2014MNRAS.441.2361V}. This reinforces the applicability of the model, at least to the levels of magnetic activity considered here. In addition, there appears to be a trend towards a steeper relation with increasing energy---from $\propto\Phi^{0.5 }$ in the EUV, to $\propto\Phi^{1.06\,- \,1.52}$ in the X-ray and SXR ranges. In agreement with observations \citep{1995A&A...300..775M} Fig. \ref{fig_11} shows that the X-ray emission will match and eventually dominate the EUV emission, at higher levels of magnetic activity (and associated magnetic flux). This is also qualitatively consistent with the spectral modelling values reported by \citet{2015Icar..250..357C}, in terms of the surface fluxes, $F_{\rm X}$ and $F_{\rm EUV}$. According to their results, this should occur at activity levels slightly higher than the ones displayed by HD 22049. This behaviour may be connected with the appearance of strong azimuthal/toroidal fields in the large-scale field of these stars, as in the case of HD 1237 (see \citealt{2015A&A...582A..38A} and references therein). 

% From Sanz-Forcada et al. (2013)              (L_X)                			(L_EUV)
%HD 1237                					8.91e+28             			5.01e+29
%HD 147513            					7.94e+28             			4.47e+29
%Eps Eri                   				1.58e+28             			2.75e+28 

% Simulations         				ZDI			SH-ZDI           	ZDI			SH-ZDI
%HD 1237                			 	3.7e+27		1.5e+28		8.5e+27		1.9e+28
%HD 147513            							7.1e+27					1.5e+28
%Eps Eri                   			2.1e+27		1.0e+28	 	5.6e+27		1.5e+28

Comparing the synthetic flux results with observations and spectral modelling data (within the same energy ranges), published by \citet{2011A&A...532A...6S}, reveals underestimated values in our models for $L_{\rm X}$ and $L_{\rm EUV}$. At best the discrepancy is less than a factor of 2 in both energy bands, as in the case of the SH-ZDI solution of HD 22049. However these differences can range up to $\sim$\,$1 - 2$ orders of magnitude, in some of our other stellar simulations. The largest discrepancies appear in the EUV band, reflecting the model emission measure deficiencies noted previously. Several observational and numerical factors could give rise to the relatively large mismatch of the results. These include instrumental and S/N effects in the EUV/X-ray observations (see \citealt{2011A&A...532A...6S}), the spatial resolution and missing flux in the ZDI reconstruction (see \citealt{2011MNRAS.410.2472A,2014MNRAS.439.2122L}), temporal incoherence (connected with long-term variations associated with magnetic cycles), the coronal heating model assumed, among others. Previous numerical studies have adjusted the thermodynamic base conditions to match the peak of the observed $EM(T)$ distribution \citep{2012MNRAS.423.3285V}, or the X-ray luminosity \citep{2013MNRAS.436.2179L}. However, as the dominant coronal emission changes with the magnetic activity of the star \citep{1995A&A...300..775M, 2015Icar..250..357C}, this has to be performed in all high-energy bands for a consistent calibration. Despite the various problems and limitations, these comparisons serve as a benchmark to improve this data-driven approach and make it more reliable in stars different from the Sun.

Finally while both solar cases agree well with observed mean coronal temperatures and high-energy emission (see Sect. \ref{SC_Calibration}), the activity maximum solution appears far up from the general trends in both the EUV and X-ray as shown in Fig. \ref{fig_11}. By removing these points from the power-law fits, the scatter is reduced considerably (by a factor of $\sim\,$5 in the EUV and $\sim\,$2.5 in the X-ray). This is directly connected with the resulting shape of the $EM(T)$ distribution, presented in Sect. \ref{cor_prop}, which is much more flatter than observed. Consequently, the EUV flux is much larger by a commensurate margin, compared to the solar minimum and the stellar counterparts. The ``cool'' plasma excess is related to the response of the coronal model heating law to the field topology and associated complexity (see Sect.~\ref{sec_stars}). To show this in more detail, we compare the surface distribution of $\Phi_{\rm Br}$ of four of the cases considered in ascending order of $\left<\Phi_{\rm Br}\right>_{\rm s}$ (Fig.~\ref{fig_12}). Even with a degraded resolution, the solar maximum case contains a far more complex field distribution than in any of our ZDI models. This makes the comparison between the solar activity maximum case and ZDI-driven models extremely difficult. From this perspective, the solar minimum state provides a more suitable point of comparison for ZDI-based stellar studies. Indeed, the predicted fluxes for the solar minimum case are well-aligned with the power-law fits in Fig.~\ref{fig_11}.

\subsection{Coronal Features and Rotational Modulation}

\noindent In this last section we calculate the rotational modulation of the high-energy emission due to the specific coronal features developed in the simulation. The stellar emission in the X-ray and EUV ranges play a fundamental role in the thermal structure and dynamical evolution of planetary atmospheres (see \citealt{2003ApJ...598L.121L,2013oepa.book.....L}). Processes like heating of the exospheres/thermospheres, expansion, and atmospheric escape are highly sensitive to these parameters \citep{2008SSRv..139..399L, 2011ApJ...733...98G, 2014ApJ...795..132S}. However, there are several observational difficulties, particularly in the EUV, in accessing these ranges of the electromagnetic spectrum, such as lack of instrumentation and strong absorption by the interstellar medium \citep{2015Icar..250..357C}. Various alternatives have been used to overcome these issues, including extrapolations based on average solar EUV fluxes \citep{2007A&A...461.1185L}, coronal models from spectral synthesis \citep{2011A&A...532A...6S}, and predictions from rotational evolution models \citep{2015A&A...577L...3T}. Still, these procedures are not able to estimate the variability in time-scales comparable to the stellar rotation period, or from the geometrical configuration of the system (e.g. orbital inclination). Both elements can be considered in our data-driven numerical approach, provided that the entire three-dimensional structure of the corona is generated. These factors can have important effects on exoplanetary conditions such as climate patterns and habitability (e.g. \citealt{2014RSPTA.37230084F}), and in the detectability of transits either in X-ray \citep{2013ApJ...773...62P}, or near UV wavelengths \citep{2012ApJ...760...79H, 2011MNRAS.416L..41L}. 

\begin{figure}[h]
\centering %  left, bottom, right and top
\includegraphics[trim=2.1cm 0.0cm 0.0cm 1.0cm, clip=true, width=\hsize]{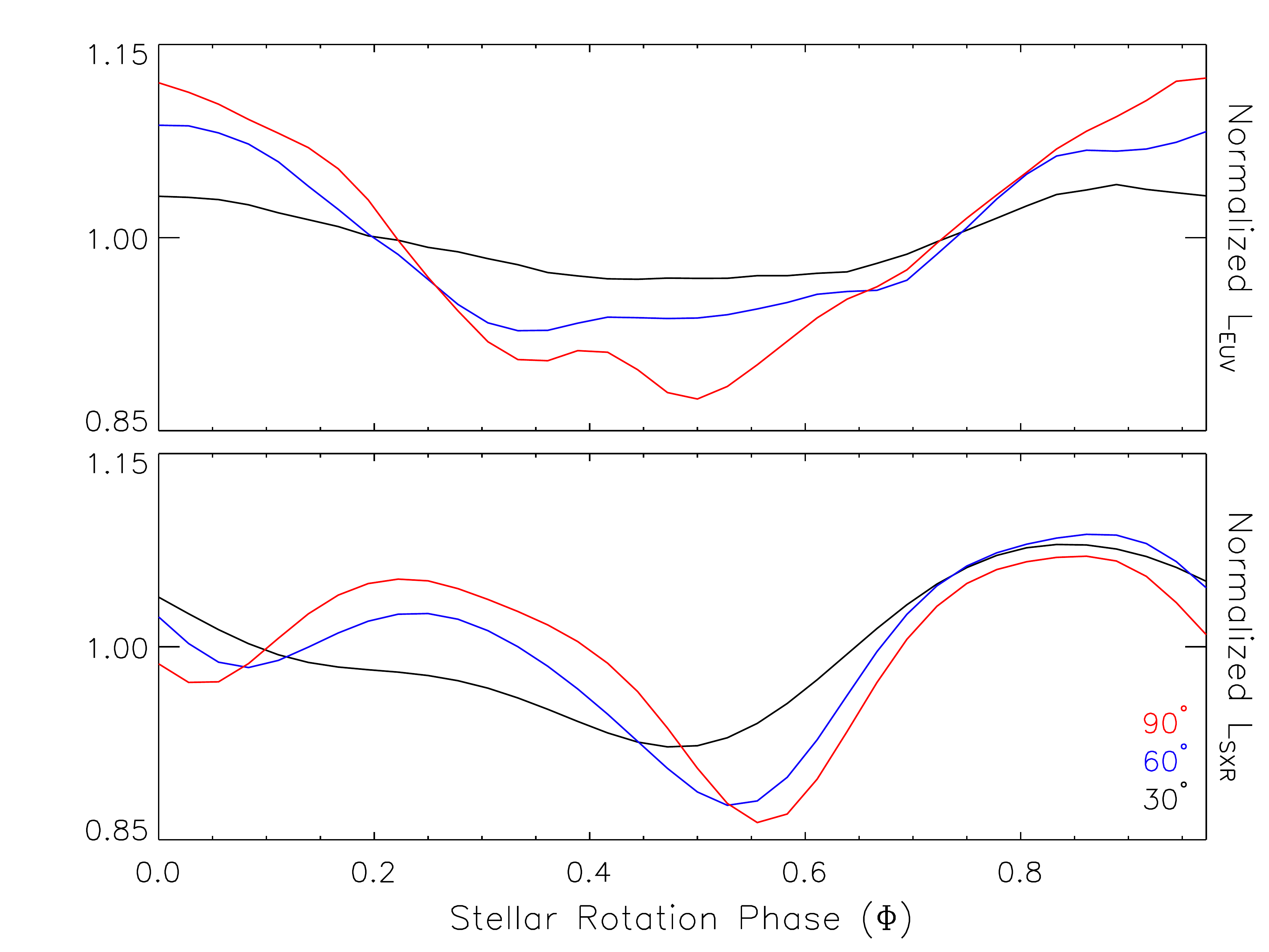}
\caption{Rotational modulation of the high-energy coronal emission for HD 1237 (SH-ZDI), showing the EUV (top) and SXR (bottom) ranges. Colours indicate the inclination angle used for the calculation.}
\label{fig_13}
\vspace{-9pt}
\end{figure}

For this purpose, we have used the most extreme system analysed here, in terms of high-energy emission and proximity of the planet. This corresponds to HD~1237 using the SH-ZDI field distribution (Fig. \ref{fig_6}, see also Table \ref{table_1}). From the steady-state coronal solution, we generate a set of synthetic high-energy emission maps covering an entire rotation of the star, with three different line-of-sight angles (30, 60 and 90$^{\circ}$)\footnote[2]{This angle is measured between the stellar rotation axis and the position of the observer.}. Figure \ref{fig_13} contains the resulting rotational modulation of the coronal emission, for each of the considered inclinations. These variations are induced by the different coronal features, described in Sect. \ref{SC_HD1237}. Around 0.2 and 0.8 in rotational phase, the large coronal holes are crossing the stellar disk, while at $\Phi = 0.5$ and $\Phi = 1.0$, they are located near the limb (close to the perspective presented in Fig. \ref{fig_6}). 

As expected, the high-energy modulation is reduced for smaller inclination angles (e.g. closer to a pole-on view), being less than 5\% of the mean value for both energy bands. For larger inclinations, the modulation increases, reaching up to $\sim$\,15\% in the 90$^{\circ}$ inclination case. These values are fully consistent with the X-ray modulation estimates in the HD 189733 (K2V) system, obtained by \cite{2013MNRAS.436.2179L}, which displays a comparable field strength at the stellar surface ($\sim$\,$\pm 30$ G). %However, the numerical model used by \citet{2013MNRAS.436.2179L} considers a coronal base temperature of $2 \times 10^{6}$ K, making the predictions for the EUV range more difficult (see Sect. \ref{cor_prop}).
 
Depending on the magnetic field evolution, the modulation of the coronal emission could persist for time-scales longer than one rotation period of the star. This is the case for HD 1237, where the large-scale field seems stable on a time-scale of months \citep{2015A&A...582A..38A}. A fraction of the coronal may show rotationally modulated coronal emission, on time-scales comparable to the orbital period of the exoplanet ($P_{\rm orb} = 133.7 \pm 0.2$ days, \citealt{2001A&A...375..205N}). However, an additional component will be due to shorter term reconnection events (e.g. flares). Furthermore, as the planetary system has a high orbital eccentricity ($e \simeq$\, 0.5), any X-ray and EUV rotational modulation will likely play a secondary role in terms of the irradiation environment. This will be considered in the second paper of this study, in combination with the influence of the magnetized stellar wind on the exoplanetary conditions.

A similar approach can be followed in other systems, provided ZDI reconstructions are available for the host star (see \citealt{2013MNRAS.435.1451F}). More complex scenarios are expected in hot-Jupiters, where the orbital period can be equal to or even shorter than the stellar rotation period. One example is the HD 179949 (F8V) system \citep{2012MNRAS.423.1006F}, where planet-induced coronal activity has been suggested \citep{2008ApJ...676..628S}. Still, recent multi-wavelength observations of this system, presented by \cite{2013A&A...552A...7S}, appear to be consistent with rotational modulation alone. Nevertheless, this does not exclude the possibility of star-planet interactions in the system, which could be explored in more detail with the data-driven approach presented in this paper. Moreover, as has been shown by \citet{2011ApJ...738..166C}, it is also possible to simulate in detail space weather phenomena, such as coronal mass ejections, in these extreme exoplanetary systems.

\section{Summary and Conclusions}\label{sec_conclusions}

\noindent We have performed a detailed numerical simulation of the 3D coronal structure of three late-type planet-hosting stars (HD 1237, HD 22049 and HD 147513). A steady-state solution is self-consistently calculated, driven by the surface magnetic field distributions recovered with the technique of Zeeman Doppler Imaging. The main results of our study are summarised below.
\begin{list}{$\circ$}{}
\item We compared the coronal solutions driven by two similar implementations of this mapping technique (ZDI and SH-ZDI). The global structure of the resulting corona is consistent in both cases. A quantitative analysis showed important differences in the thermodynamic conditions, and in the coronal high-energy emission. We obtain differences up to factors of $1.4$ and $2.5$ in the coronal temperature and density, respectively. This led to a larger variation in the predicted EUV and SXR emission, reaching up to one order of magnitude. These differences can be related to the amount of structure, field strength, and the map completeness in each case. \vspace{2pt}
 
\item The appearance of different coronal features in each star is highly dependent on the characteristics of the surface field distribution. In the case of HD 1237, two large coronal holes appear as the most prominent elements. HD 22049 shows more complex details, displaying additional structures such as helmet streamers and filaments. For HD 147513, the simulation predicts a rather simple coronal topology, reflecting the low-complexity of its surface magnetic field.\vspace{2pt}
 
\item Comparable solar simulations, in terms of spatial resolution and boundary conditions, were considered (covering activity minimum and maximum). This included a detailed comparison with archival satellite data in the EUV and SXR ranges. For both activity states, good agreement was obtained in terms of the coronal temperature ($\sim$ 8\% difference), and in the high-energy coronal emission (SXR/EUV bands). On the other hand, the emission measure distribution showed larger discrepancies, with a considerable excess in the low-temperature end ($\log T < 6.0$) for the solar maximum case. In addition, within the temperature range of $1 - 2$ MK, the EM appears underestimated by factors of $\sim$\,3 and 5, for activity maximum and minimum, respectively. This is likely indicative of the need to recalibrate the coronal heating mechanism, when applying this model to resolution-limited surface magnetic field distributions. \vspace{2pt}
  
\item Furthermore, while the comparison to the observations showed similar levels of agreement for both solar minimum and maximum cases (e.g. thermodynamic conditions and high-energy emission), the simpler structure of the large-scale magnetic field makes the former a better reference point for simulations based on ZDI maps (see Fig. \ref{fig_12}). \vspace{2pt}

\item We considered the particular case of HD 1237, to estimate the rotational modulation in the high-energy emission due to the coronal features developed in our simulation. We obtain variability ranging from $\sim$\,$5 - 15$\% (depending on the line-of-sight angle), in the mean coronal EUV and SXR emission. Similar estimates have been reported for systems with comparable surface field strengths (e.g. HD 189733, \citealt{2013MNRAS.436.2179L}). \vspace{2pt} 
 
\item In addition, using the simulations we were able to recover similar trends as in previous observational studies, including a relation between the magnetic flux ($\Phi_{\rm Br}$) and the coronal high-energy emission \citep{2003ApJ...598.1387P, 2014MNRAS.441.2361V}. However, as this numerical model was specifically developed for the Sun, further adjustments will be required to better calibrate our results to the stellar data. \vspace{2pt}

\item Improvements in this approach can be performed by extending the range in $\Phi_{\rm Br}$. This could be done by isolating specific regions from high-resolution solar observations, and by expanding the stellar sample to more active stars. For the latter case, it would be necessary to further adjust the coronal heating or the thermodynamic base conditions, to match the observed coronal emission in all energy bands. Another possibility would involve a more sophisticated numerical treatment, in order to consider all the magnetic field components to drive the simulation (see \citealt{2015SpWea..13..369F}). \vspace{2pt}

\item The results discussed in this work will be used, in a follow-up paper, to self-consistently simulate the stellar wind, inner astrosphere and circumstellar environment of these systems. This includes stellar and planetary mass losses, orbital conditions and topology of the astrospheric current sheet.

\end{list}

\begin{acknowledgements}
\noindent This work was carried out using the SWMF/BATSRUS tools developed at The University of Michigan Center for Space Environment Modeling (CSEM) and made available through the NASA Community Coordinated Modeling Center (CCMC). We acknowledge the support by the DFG Cluster of Excellence "Origin and Structure of the Universe". We are grateful for the support by A. Krukau through the Computational Center for Particle and Astrophysics (C2PAP).
\end{acknowledgements}

\bibliographystyle{aa}
\bibliography{Biblio}

\begin{appendix}

\begin{figure*}[ht]
\section{Simulation Results for the Sun}\label{app_1}
\centering%  left, bottom, right and top
\includegraphics[scale=0.333]{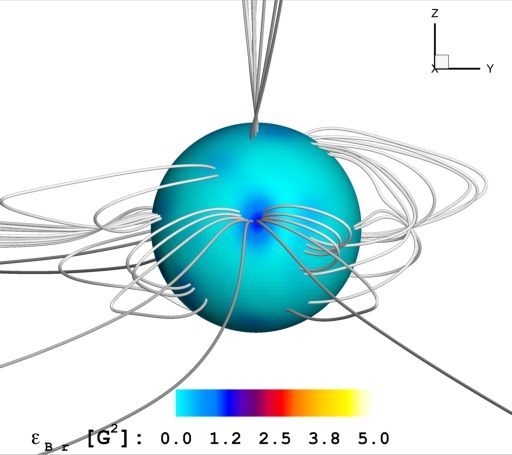}\hspace{2pt}\includegraphics[scale=0.333]{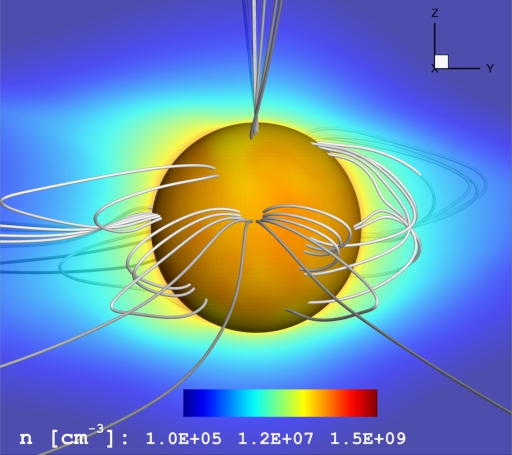}\hspace{2pt}\includegraphics[scale=0.333]{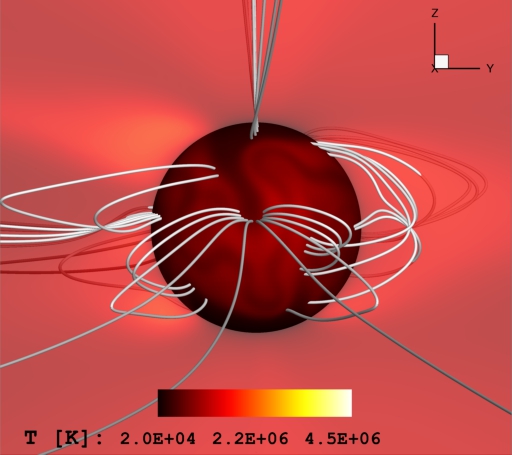}\vspace{1pt}
\includegraphics[scale=0.22505]{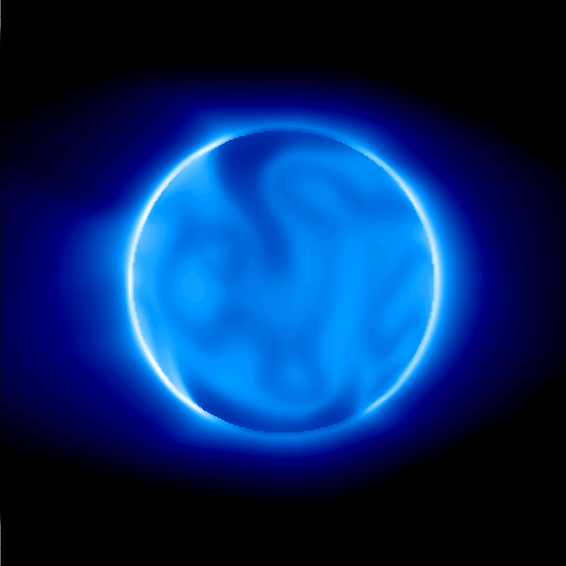}\hspace{2pt}\includegraphics[scale=0.22505]{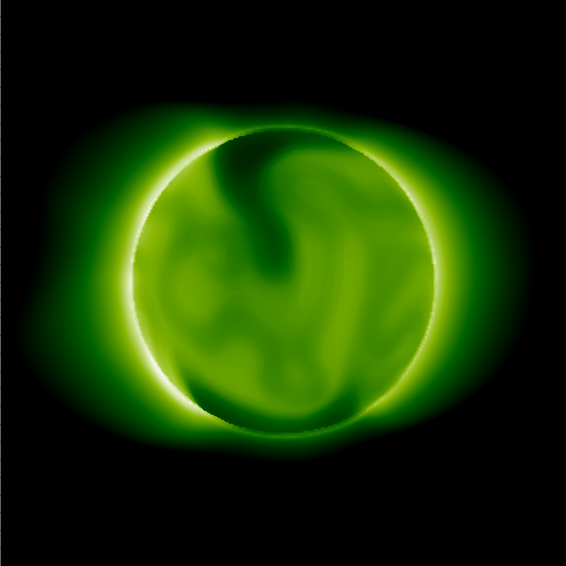}\hspace{2pt}\includegraphics[scale=0.22505]{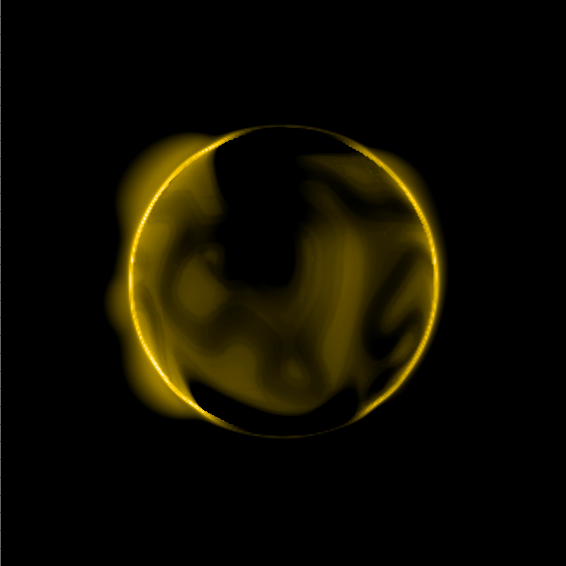}\hspace{2pt}\includegraphics[scale=0.22505]{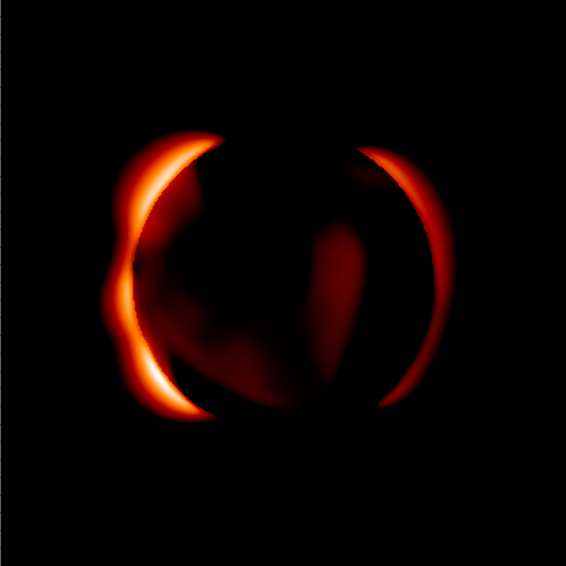}
\caption{Simulation results for the coronal structure of the Sun during activity minimum (CR 1922). The upper panels contain the distribution of the magnetic energy density ($\varepsilon_{\,B\,r}$, left), the number density ($n$, middle) and temperature ($T$, right). For the last two quantities, the distribution over the plane $y = 0$ is presented. The sphere represents the stellar surface and selected three-dimensional magnetic field lines are shown in white. Note the change of scale for $\varepsilon_{\,B\,r}$ in this case. The lower images correspond to synthetic coronal emission maps in EUV (blue / 171\,\AA\,, green / 195\,\AA\, and yellow / 284\,\AA) and SXR (red / 2 -- 30\,\AA). The perspective is preserved in all panels, with an inclination angle of $i = 90\,^{\circ}$.}
\label{fig_A1}
%\end{figure*}
\vspace{10pt}
%\begin{figure*}[!h]
\centering%  left, bottom, right and top
\includegraphics[scale=0.333]{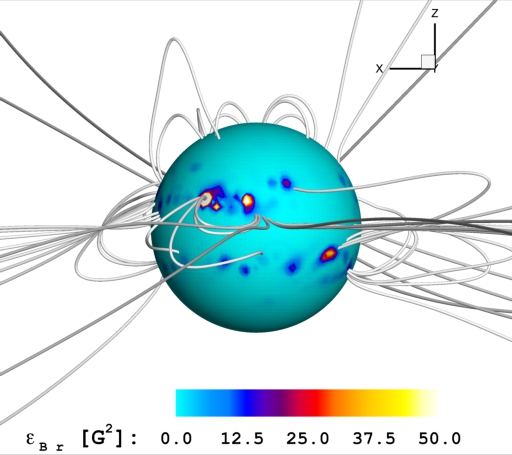}\hspace{2pt}\includegraphics[scale=0.333]{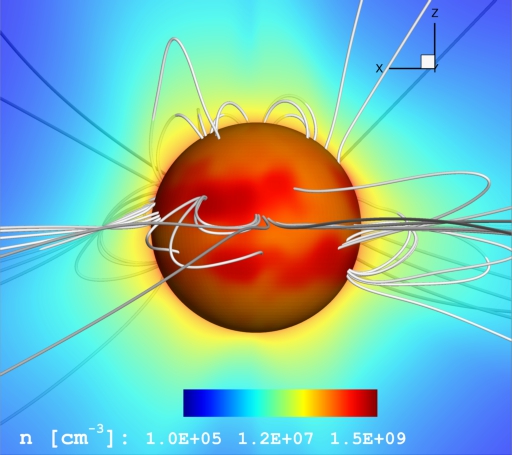}\hspace{2pt}\includegraphics[scale=0.333]{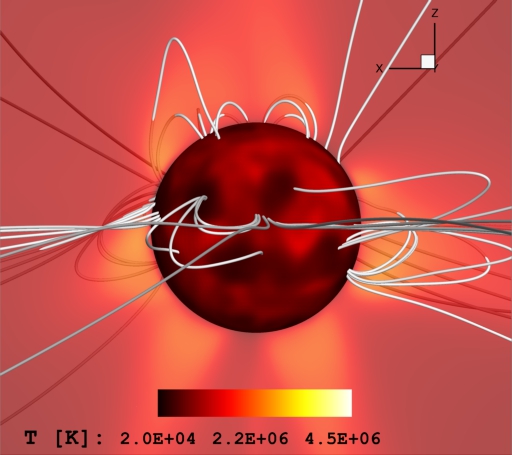}\vspace{1pt}
\includegraphics[scale=0.22505]{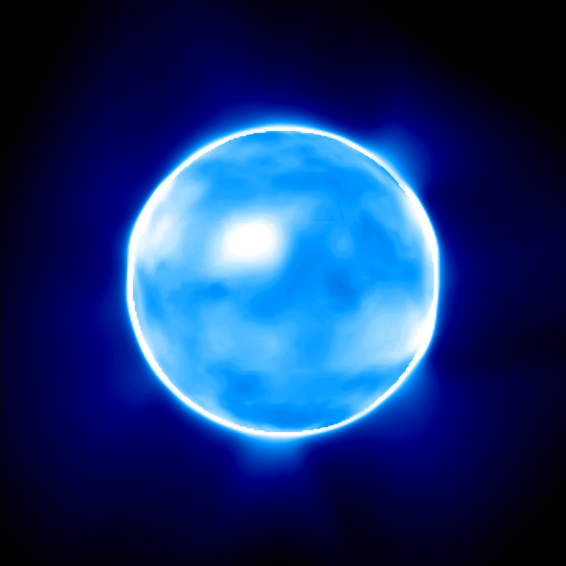}\hspace{2pt}\includegraphics[scale=0.22505]{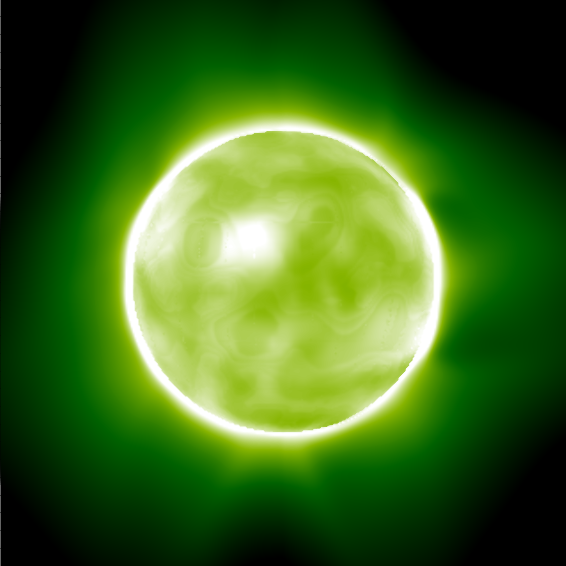}\hspace{2pt}\includegraphics[scale=0.22505]{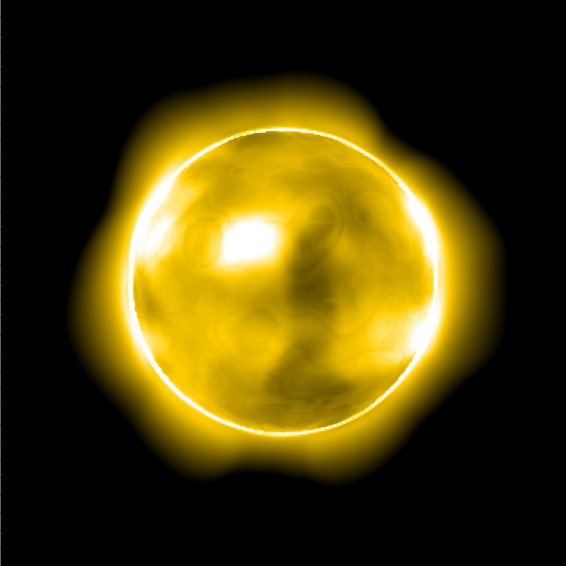}\hspace{2pt}\includegraphics[scale=0.22505]{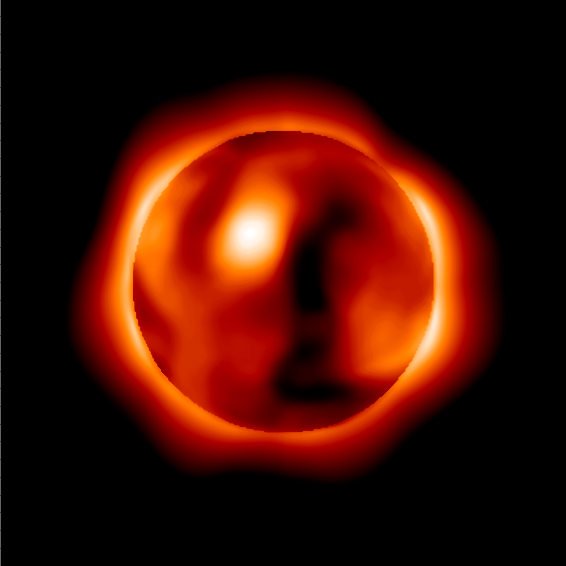}
\caption{Simulation results for the coronal structure of the Sun during activity maximum (CR 1962). See caption of Fig. \ref{fig_A1}.}
\label{fig_A2}
\end{figure*}

\end{appendix}

\end{document}